\documentclass[11pt]{article}
\usepackage{amsmath,amssymb}

\newsavebox{\uuunit}
\sbox{\uuunit}
    {\setlength{\unitlength}{0.825em}
     \begin{picture}(0.6,0.7)
        \thinlines
        \put(0,0){\line(1,0){0.5}}
        \put(0.15,0){\line(0,1){0.7}}
        \put(0.35,0){\line(0,1){0.8}}
       \multiput(0.3,0.8)(-0.04,-0.02){10}{\rule{0.5pt}{0.5pt}}
     \end {picture}}

\long\def\symbolfootnote[#1]#2{\begingroup%
\def\thefootnote{\fnsymbol{footnote}}\footnote[#1]{#2}\endgroup}
\numberwithin{equation}{section}

\begin{document}
\begin{titlepage}
\begin{center}
\hfill DFPD-09TH09\\
\hfill LMU-ASC 25/09 \\
\hfill MPP-2009-67
\vskip 10mm

{\LARGE \textbf{On Subextensive Corrections to Fluid Dynamics
\\ [3mm]
from Gravity
}}
\vskip 12mm

\textbf{G.L.~Cardoso$^{a}$, G.~Dall'Agata$^{b} $ and V.~Grass$^{c,d}$} 

\vskip 6mm
$^a${\em CAMGSD, Departamento de Matem\'atica\\
Instituto Superior T\'ecnico, Av. Rovisco Pais 1, 1049--001 Lisboa, Portugal}\\
[3mm]
$^b${\em Dipartimento di Fisica ``Galileo Galilei'' $\&$ INFN, Sezione di
  Padova \\
Universit\`a di Padova, Via Marzolo 8, 35131 Padova, Italy}\\
[3mm]
$^c${\em Max-Planck-Institut f\"ur Physik\\
F\"ohringer Ring 6, 80805 M\"unchen, Germany}\\
[3mm]
$^d${\em Arnold Sommerfeld Center for Theoretical Physics, 
Department f\"ur Physik\\
Ludwig-Maximilians-Universit\"at M\"unchen, 
Theresienstr. 37, 80333 M\"unchen, Germany}
\end{center}
\vskip .2in
\begin{center} {\bf ABSTRACT } \end{center}
\begin{quotation}\noindent
We use the fluid-gravity correspondence to compute subextensive
corrections, proportional to the shear tensor, to the 
energy-momentum tensor of fluids on three-spheres.
The dual configurations we consider are charged black hole solutions of $N = 2$ gauged supergravity theories in five dimensions.

\end{quotation}

\vfill
\end{titlepage}
\eject

\section{Introduction}

The conformal fluid-gravity correspondence relates the hydrodynamic regime
of strongly coupled four-dimensional conformal field theories 
to regular black brane
solutions in asymptotically $AdS_5$ backgrounds
\cite{Policastro:2001yc,Policastro:2002se}.
The black brane 
solutions are constructed order by order in a gradient expansion
in the bulk, and this gradient expansion is mapped to the hydrodynamic
gradient expansion 
of the fluid's energy-momentum tensor $T_{\mu \nu}$
in the dual boundary theory \cite{Bhattacharyya:2008jc}.
The coefficients in the gradient expansion of $T_{\mu \nu}$
are the hydrodynamic transport 
coefficients that characterize the hydrodynamic properties of the fluid,
and these are holographically determined in terms of the black brane solutions.

The gradient expansion of the fluid's energy-momentum tensor contains
a term proportional to the shear tensor $\sigma_{\mu \nu}$, with a 
coefficient 
$\eta$
that has been computed for various
conformal fluids dual to black branes, 
starting with \cite{Policastro:2001yc}.
For these fluids $\eta$ can be expressed in terms of the energy density $\rho$, the pressure $p$ and the diffusion coefficient
$D$ as \cite{Policastro:2002se}
\begin{equation}
\eta = \left( \rho + p \right) \, D \;.
\label{eq:eta-D}
\end{equation}
For uncharged conformal fluids, $D$ is expressed in terms of the entropy density $s$ of the black brane as $D = \pi^{1/3}/(16 \, s)^{1/3}$ (in units where $L = 16 \pi G_5 = 1$, where $L$ denotes the curvature radius of $AdS_5$)
\cite{Policastro:2002se}. It follows from (\ref{eq:eta-D}) that
\begin{equation}
\frac{\eta}{s} = \frac{\pi^{1/3}}{4^{2/3} } \,
\frac{ \rho + p }{ s^{4/3}} \;,
\label{eq:eta-s-e}
\end{equation}
which
equals $\eta/s= 1/(4 \pi)$, since for a conformal fluid $\rho = 3 \, p = 3 \, s^{4/3}/(4 \pi)^{4/3} $. This behavior of $\eta/s$ is also observed for charged fluids, so that for conformal (charged) fluids dual to (charged) black branes, the ratio $\eta/s$ seems to take the universal value $1/(4 \pi)$ at strong 't Hooft coupling and in the large $N$ limit \cite{Policastro:2001yc,Policastro:2002se, Bhattacharyya:2008jc,Buchel:2003tz,Buchel:2004qq, Mas:2006dy,Son:2006em,Maeda:2006by,Benincasa:2006fu,
Eling:2008af,Erdmenger:2008rm,Banerjee:2008th,Iqbal:2008by,
Torabian:2009qk}.

In this note we will focus on charged black holes 
in asymptotically $AdS_5$ backgrounds,
rather than charged black branes.  
The black hole solutions we consider 
capture the hydrodynamic expansion of the dual 
conformal fluid on a three-sphere 
\cite{Bhattacharyya:2007vs,Caldarelli:2008mv,Caldarelli:2008ze,
Cardoso:2009bv}.
In contrast to fluids in flat space,  
the energy of a fluid on a three-sphere is not anylonger 
a purely extensive quantity
\cite{Verlinde:2000wg,Savonije:2001nd}.  It contains, in particular, a subextensive
part $E_c$ which is defined as the violation
of the thermodynamic Euler relation \cite{Verlinde:2000wg}.  
One may ask whether this non-extensivity will result in
a correction of the coefficient $\eta$ of the shear tensor and hence in
a deviation from the
value $\eta/s= 1/(4 \pi)$ for these fluids 
(at strong 't Hooft coupling and in the large $N$ limit).
Consider, for instance, the conformal fluid dual to a Schwarzschild 
black hole.  Its total energy $E$ equals 
$E = E_e + \frac12 E_c$ and the associated density is $\rho = \rho_e + \frac12
\rho_c$. Here $E_e$ denotes the extensive part of the energy, and its
energy density is 
$\rho_e = 
3 \, s^{4/3}/(4 \pi)^{4/3}$. Taking   
the relation (\ref{eq:eta-s-e}), which has been derived for flat branes,
at face value then
suggests that 
\begin{equation}
\frac{\eta}{s} = \frac{1}{4 \pi} \, \left(1 +  
\frac{\rho_c}{2 \rho_e} \right)\;.
\label{eq:change-ration-schw}
\end{equation}
In this note we will show that 
(\ref{eq:change-ration-schw}) indeed holds for 
the conformal fluid dual
to a Schwarzschild black hole. 
For a discussion of a similar effect for 
fluids on hyperbolic spaces see \cite{Koutsoumbas:2008wy}.
In the charged case, the black holes that we consider
arise in the so-called STU-model of $N=2$ gauged supergravity
in five dimensions. 
We use the formalism developed in 
\cite{Bhattacharyya:2007vs,Bhattacharyya:2008jc,Loganayagam:2008is,
Erdmenger:2008rm,Banerjee:2008th,Bhattacharyya:2008mz}
to construct these electrically charged deformed 
black hole solutions.
We again observe a deviation 
from the value $\eta/s= 1/(4 \pi)$ in all these cases.

This deviation can be understood as follows.
The relation \eqref{eq:eta-s-e}
 was established at first order in the derivative expansion.  At this order, $\eta$ has the hydrodynamical interpretation of shear viscosity, since it denotes the coefficient of the shear tensor $\sigma_{\mu \nu}$ in the gradient expansion of $T_{\mu \nu}$. At higher order, however, the fluid's energy-momentum tensor may contain
additional higher-derivative terms that are also proportional to the shear tensor.  For instance, at cubic order in
derivatives,  there may be an additional term of the form $R \, \sigma_{\mu \nu}$, where
$R$ denotes the curvature scalar of the three-sphere on which the dual fluid lives.
Then, combining all the terms proportional to the shear tensor in $T_{\mu \nu}$, yields a shear term with an effective coefficient
$\eta$ that will exhibit a departure from the first-order value $s/(4 \pi)$.  Whether or not 
this effective coefficient $\eta$ 
continues to satisfy relation  \eqref{eq:eta-s-e} is, a priori, not known.

In \cite{Bhattacharyya:2007vs} it was shown that large rotating black holes in global $AdS_D$ spaces are
dual to stationary solutions of the relativistic Navier-Stokes equations on $S^{D-2}$.
The dual description in terms of fluid dynamics applies when
various length scales, namely the one associated with the curvature of the manifold on which the fluid propagates and those describing the variation of the thermodynamic variables,
are large compared to the equilibration length scale of the fluid. As shown in \cite{Bhattacharyya:2007vs},
this requires taking the horizon radius $R_H$ of the dual black hole to be large compared to
the AdS radius $R_{AdS}$.  These black holes are 
non-supersymmetric and are 
referred to as large black holes.
Then, expanding the black hole formulae in a power series in
$R_{AdS}/R_H$ results in subleading corrections that show up as corrections in the
energy-momentum tensor of the dual fluid.  The same continues to hold when
considering non-stationary black holes.  For large black holes, the subleading corrections in $R_{AdS}/R_H$ in the black hole formulae will contribute to the gradient expansion of the fluid's energy-momentum tensor. An example thereof is the term proportional to $\rho_c/\rho_e$ in 
\eqref{eq:change-ration-schw}.  For large black holes it constitutes a small correction to the ratio $\eta/s$.

This note is organized as follows.  In section \ref{elec-bhs-static} we 
review electrically
charged static black hole solutions with spherical horizons
of certain 
five-dimensional $N=2$ gauged supergravity theories.
In section \ref{def-sol-stu} we deform these solutions by a slowly
varying velocity field and we explain our procedure for
determining corrections to $\eta/s$ induced by the curvature $k$ of
the fluid's three-sphere. Then we turn to black holes (with up to three 
equal charges) in
the context of the STU-model, and
we compute the first correction in $k$ to $\eta/s = 1/(4 \pi)$. 
Section \ref{concl} contains our conclusions.  Appendix \ref{sec:VSG} summarizes our very special
geometry conventions.  For the sake of comparison with the deformed solutions
in section 3, we summarize various known rotating solutions of the STU-model
in appendix \ref{bl-ef}, \ref{STUblackhole} and \ref{genSTU}.  And finally, appendix \ref{tmnSTU} summarizes the calculation
of the boundary energy-momemtum tensor for one of the black hole solutions of the
STU-model.

\section{Electrically charged static black hole solutions
\label{elec-bhs-static}}

We begin by reviewing the electrically charged static 
black hole solutions constructed in \cite{Behrndt:1998jd}.
These will subsequently be deformed by a non-trivial velocity field.
The static solutions of \cite{Behrndt:1998jd} are solutions
of five-dimensional $N=2$ gauged supergravity theories 
obtained by gauging 
the $U(1)$ subgroup of the $SU(2)$-automorphism group of
the $N=2$ supersymmetry algebra \cite{Gunaydin:1984ak}.  The gauging is 
with respect to a linear combination 
proportional to $h_A \, A_M^A$ 
of $U(1)$ gauge fields
(with constant $h_A$), 
and the coupling constant $\mathfrak{g}$ is identified with the
inverse of the curvature radius of $AdS_5$, i.e. $\mathfrak{g} = L^{-1}$.
The relevant part of the action reads \cite{Gunaydin:1984ak}
\begin{eqnarray}
16 \pi G_5\; S &=& \int d^5 x \sqrt{-G} \left( R -
 {\cal G}_{ij} \, \partial_M \varphi^i
\, \partial^M \varphi^j
- \frac12 \, {\cal G}_{AB} F^A_{MN} \, F^{B MN} - V_{\rm pot}
\right) \nonumber\\
&&
+ \frac{\kappa}{\sqrt{3}} \,
\int C_{ABC} \, F^A \wedge F^B \wedge A^C \;,
\label{action5}
\end{eqnarray}
where $\kappa = -1/(2 \sqrt{3})$.
We denote the five-dimensional spacetime metric by $G_{MN}$.
We refer to appendix \ref{sec:VSG} for a definition of the various quantities
appearing in (\ref{action5}).

The static charged black hole solutions we consider
are black holes with a spherical horizon.  
Their line element reads \cite{Behrndt:1998jd} 
\begin{equation}
ds^2 = G_{MN} \, dx^M dx^N = - {\rm e}^{-4U(r)} \, p(r) \, dt^2 + 
{\rm e}^{2U(r)} \, {p}^{-1}(r) \, dr^2
+ {\rm e}^{2U(r)} \, r^2 \, d\Omega_3^2 \;,
\label{eq:line-static}
\end{equation}
where 
\begin{equation}
p(r) = 
k - \frac{\mu}{r^2} + \frac{{\rm e}^{6U}\,r^2}{L^2} \;\;\;,\;\;\; k> 0\;.
\end{equation}
The line element of the three-sphere can be written as
\begin{equation}
 d\Omega_3^2 = g_{ij} \, dx^i dx^j = k^{-1} \left(
d \theta^2 + \sin^2 {\theta} \, d \phi^2
+ \cos^2 \theta \, d \psi^2 \right) \;,
\label{eq:bound-metric}
\end{equation}
with $0 \leq \theta \leq \pi/2 \, , \, 0 \leq \phi < 2 \pi \, , \,
 0 \leq \psi < 2 \pi$. The curvature tensor of the three-sphere is
$R_{ij} = 2 \, k \, g_{ij}$, and the associated curvature scalar is 
$R = 6 \,k$.  

These black hole solutions are supported by scalar fields $X^A (r)$.
They satisfy the relation
\begin{equation}
X_A = \frac13 \, {\rm e}^{-2U} \, H_A \;,
\end{equation}
where the $H_A$ denote harmonic functions given by $H_A = 
h_A + q_A/r^2$. The parameters $q_A$ are related to the electric charges
and to the mass of the black hole solutions, as we will discuss
below. The metric factor ${\rm e}^{2 U}$ is given by 
\begin{equation}
{\rm e}^{2 U} = \frac13 \, H_A \, X^A \;, 
\end{equation}
and its radial derivative $U' = d U/dr$ is related to the 
superpotential $W = h_A X^A$ by
\cite{Cardoso:2008gm},
\begin{equation}
{\rm e}^{2 U(r)} \left( 1 + r \, U' \right) = \frac{W}{3} \;.
\label{eq:W-Uprime}
\end{equation}

We take $h_A$ and $q_A$ to be positive to ensure that 
$H_A >0$.  We also take $X^A >0$ so that ${\rm e}^{2 U} > 0$.
We impose the normalization ${\rm e}^{2 U} = 1$ at $r = \infty$.
The asymptotic value of $X_A$ is then $\frac13 h_A$.
Denoting 
the asymptotic value of the $X^A$ by $h^A$, we have $ \frac13 h^A \, h_A =1$
in view of real special geometry (see
\eqref{vsgconstraint}).  Using $h^A$, 
we introduce the `dual'
superpotential
${\tilde W}$ as 
\begin{equation}
{\tilde W} = h^A \, X_A \;,
\label{dualsupo}
\end{equation} 
for later convenience \cite{Cardoso:2008gm}.  It asymptotes to ${\tilde W}=1$, while the superpotential $W$ 
asymptotes to the value $W=3$.

The mass $M$ of the black hole and its 
physical electric charges $Q_A$ are determined in terms of the parameters
$\mu$ and $q_A$ as follows \cite{Cardoso:2008gm},
\begin{eqnarray}
w_5 \, M \,  &=& \mu + \frac23 \, k \, h^A \, q_A \;, \nonumber\\
Q_A G^{AB} Q_B &=& k \, q_A G^{AB} q_B + \mu \, q_A G^{AB} h_B \;,
\label{eq:m-Q}
\end{eqnarray}
where $w_5 = 16 \pi \, G_5 / (3 \, vol(S^3))$, with $vol(S^3)=
\int d \Omega_3$.

Inspection of the line element (\ref{eq:line-static}) shows that the
radius of the three-sphere is ${\rm e}^U \, r$
in units of $1/\sqrt{k}$. It is thus convenient to
introduce a new radial coordinate 
$a = {\rm e}^U \, r$.  We also introduce the function 
\begin{equation}
f = {\rm e}^{-4 U} \, \frac{p}{a^2} = 
\frac{1}{L^2} + {\rm e}^{-4 U} \,\frac{k}{a^2}
- {\rm e}^{-2U} \, \frac{\mu}{a^4} \;.
\label{eq:function-f}
\end{equation}
Then, using (\ref{eq:W-Uprime}), the line element takes the form
\begin{equation}
ds^2 = - a^2 \, f(a)  \, dt^2 + 9
\left(a^2 \, f(a) \, W^2(a) \right)^{-1} da^2
+ a^2 \, d\Omega_3^2 \;.
\label{eq:line-orig}
\end{equation}
Next, we introduce 
Eddington--Finkelstein type coordinates by
\begin{equation}
v = t + g(a) \;\;\;,\;\;\; \frac{dg}{da} = \frac{3}{W(a) \, a^2 \, f(a)} 
\;,
\end{equation}
so that
the line element (\ref{eq:line-orig}) becomes
\begin{equation}
ds^2 = - a^2 \, f(a) \, dv^2 + \frac{6}{W(a)} \,
dv \, da + a^2  \, d\Omega_3^2 \;.
\label{eq:line-finkel}
\end{equation}

Following \cite{Bhattacharyya:2007vs,Bhattacharyya:2008jc}, 
we define boundary coordinates $x^{\mu} = (v, \theta, \phi, \psi)$ and
we introduce the associated 
four-dimensional metric $g_{\mu \nu} = (g_{vv}, g_{ij}) = 
(-1, g_{ij})$, which will be kept fixed throughout.
Then, the static black hole metric (\ref{eq:line-finkel}) can be written
as
\begin{equation}
ds^2 = - a^2 \, f(a) \, u_{\mu} \, u_{\nu} \, dx^{\mu} \, dx^{\nu}
- \frac{6}{W(a)} 
\, u_{\mu} \, dx^{\mu} \, da + a^2  \, P_{\mu \nu} \, dx^{\mu} \, 
dx^{\nu} \;,
\label{eq:line-boost1}
\end{equation}
where here $u_{\mu}$ denotes the four-vector $u_{\mu} = (-1,0,0,0)$ and
where
\begin{equation}
P_{\mu \nu} = g_{\mu \nu} + u_{\mu} \, u_{\nu} \;.
\end{equation}
The four-vector $u_{\mu}$ denotes the velocity vector of
the dual fluid.
Indices of boundary tensor quantities will be lowered or raised
using the boundary metric $g_{\mu \nu}$ and its inverse $g^{\mu \nu}$, such as, for instance, 
$u^{\mu} = g^{\mu \nu} \, u_{\nu}$.

In the following, we set $L=1$ for convenience.
Following \cite{Loganayagam:2008is,Bhattacharyya:2008mz}, 
we introduce the Schouten tensor
$S_{\mu \nu} = \frac12 \left(R_{\mu \nu} - \frac16 \, R \, g_{\mu \nu}
\right)$.
Here $R_{\mu \nu}$ and $R$ are the four-dimensional Ricci tensor and 
Ricci scalar 
computed from the metric $g_{\mu \nu}$.  
Then, 
the line element (\ref{eq:line-boost1}) can also be expressed as
\begin{equation}
ds^2 = 
- \frac{6}{W(a)} 
\, u_{\mu} \, dx^{\mu} \, da + \left[
a^2  \, g_{\mu \nu} + {\rm e}^{- 4U} \, 
u_{(\mu} \, S_{\nu) \lambda} \, u^{\lambda} + {\rm e}^{-2 U} \frac{\mu}{a^2}
\, u_{\mu} \, u_{\nu}
\right]
dx^{\mu} \, 
dx^{\nu} \;,
\label{eq:line-schout}
\end{equation}
where $a_{(\mu} \, b_{\nu)} = a_{\mu} \, b_{\nu} + a_{\nu} \, b_{\mu}$.
Observe that (\ref{eq:line-schout}) is invariant under the global 
rescaling \cite{Loganayagam:2008is,Bhattacharyya:2008mz}
\begin{equation}
\label{eq:scalings}
a \rightarrow {\rm e}^{- \chi} \, a \;\;\;,\;\;\;\;
g_{\mu \nu} \rightarrow {\rm e}^{2 \chi} \, g_{\mu \nu} \;\;\;,\;\;\;
u_{\mu} \rightarrow {\rm e}^{\chi} \, u_{\mu} \;\;\;,\;\;\; 
{\rm e}^U \rightarrow {\rm e}^U \;\;\;,\;\;\; 
\mu \rightarrow {\rm e}^{-4 \chi} \, \mu \;,
\end{equation}
which also implies the rescaling
\begin{equation}
W \rightarrow W \;\;\;,\;\;\; q_A
\rightarrow {\rm e}^{-2 \chi} \, q_A \;\;\;,\;\;\;
k \rightarrow {\rm e}^{-2 \chi} \, k \;\;\;,\;\;\; w_5 M 
\rightarrow {\rm e}^{-4 \chi} \, w_5 M \;\;\;,\;\;\; Q_A
\rightarrow {\rm e}^{-3 \chi} \, Q_A \;.
\label{eq:scalings2}
\end{equation}

Let us now discuss various black hole solutions 
in the context
of the STU-model \cite{Sabra:1997yd,Elvang:2007ba}.  
This model has three scalar fields $X^A$ that
are constrained by
$X^1 \, X^2 \, X^3 = 1$, and it allows for two solutions 
for which $W$ takes a constant value, namely
the uncharged Schwarzschild case and the charged Maxwell
black hole.  In both cases the scalar fields are constant, i.e.
$X^1 = X^2 = X^3 = 1$, and $W$ takes the value $W=3$.  
The Maxwell solution is obtained by setting
$H_1 = H_2 = H_3 = H = 1 + q/r^2$, in which case 
${\rm e}^{2U} = H$.  It follows that $a^2 = r^2 + q$ and 
${\rm e}^{-2U} = 1 -q/a^2$.
Inspection of (\ref{eq:m-Q}) yields the mass $M$ and the physical charge $Q$
as $w_5 \, M= \mu + 2 \, k \, q$ 
and $Q^2= k \, q^2 + \mu \, q$.  The Schwarzschild solution is obtained
by setting $q =0$.  In both cases the `dual' superpotential reads ${\tilde W}=1$.

The STU-model also allows for black hole solutions which are supported
by non-trivial scalar fields.  An example with two equal charges is obtained
by setting 
$H_1 = H_2 = H = 1 + q/r^2$ and $H_3 = 1$, in which case $X^1 = X^2 = H^{-1/3} \,,\,
X^3 = H^{2/3}$ as well as ${\rm e}^{3U} = H$.
Now the associated mass $M$ and physical charge $Q_1= Q_2 = Q$ 
read $w_5 \, M= \mu + \frac43 \, k \, q$ 
and $Q^2= k \, q^2 + \mu \, q$.  
On the solution, $W$ is given by $W = 2 H^{-1/3} +  H^{2/3}$.
The `dual' superpotential reads ${\tilde W} = \frac13 \left(H^{-2/3} + 2 
H^{1/3} \right)$.

An example with one non-vanishing charge is obtained by setting
$H_1 = H = 1 + q/r^2$ and $H_2 = H_3 = 1$, in which case $X^1 = H^{-2/3} \,,\,
X^2 = X^3 = H^{1/3}$ as well as ${\rm e}^{6U} = H$.
Now the associated mass $M$ and physical charge $Q_1= Q$ 
read $w_5 \, M= \mu + \frac23 \, k \, q$ 
and $Q^2= k \, q^2 + \mu \, q$.  
On the solution, $W$ is given by $W = H^{-2/3} + 2 H^{1/3}$.
The `dual' superpotential reads ${\tilde W} = \frac13 \left(H^{2/3} + 2 
H^{-1/3} \right)$.

\section{Deformed black hole solutions \label{def-sol-stu}}

In the following, we will deform the static solutions described in 
the previous section by a slowly varying
velocity field $u^{\mu}(x)$
of the form
\begin{equation}
u^{\mu} = (1, \epsilon \, \beta^{\theta}(x),
\epsilon \, \beta^{\phi}(x),
\epsilon \, \beta^{\psi}(x)) \;.
\label{eq:velo-field}
\end{equation}
Here we have multiplied the deformation $\beta$ with 
a small parameter $\epsilon$.
Thus, the deformation $u^i$ is taken to be small in amplitude.
We will work at first order in $\epsilon$.
At this order, $u^{\mu}$ satisfies 
the normalization condition $u^{\mu} \, u_{\mu} = -1$.

In addition, and following \cite{Bhattacharyya:2008jc}, 
we introduce a counting parameter $\delta$ by performing the rescaling 
$x^{\mu} \rightarrow \delta \, x^{\mu}$,  
so that an expansion in powers
of $\delta$ counts covariant derivatives.
For instance, the curvature tensor $R_{ij}$ of the three-sphere (which
we will call the background curvature tensor in the following)
will then come
multiplied by a factor $\delta^2$.

The boundary energy-momentum tensor $T_{\mu \nu}$ of the deformed
solutions contains a term proportional to the shear tensor
$\sigma_{\mu \nu}$, with a coefficient denoted by $\eta$.
We are interested in computing corrections to the ratio $\eta/s$ due to
the background curvature scalar $R= 6 k$.  
These corrections, if present, give rise to deviations from the value
$4 \pi \, \eta/s = 1$, 
which we write as
$4 \pi \, \eta/s - 1 = \sum_{p\geq 1} \alpha_{2p} \, \delta^{2p}$. 
To compute these corrections, 
we organize the perturbations of the black hole metric in powers
of $\epsilon$ and $\delta$.  In this note
we will only deal with the first 
subleading correction $\alpha_2 \, \delta^2$.  It
corresponds to a term of the type $k \, \sigma_{\mu \nu}$, and hence  of order $\epsilon \, \delta^3$, in 
the boundary energy-momentum tensor $T_{\mu \nu}$. 
Thus, we will only keep terms in the perturbed line element that 
are at most of order $\epsilon \, \delta^3$.

Let us first consider the Schwarzschild case.  The static Schwarzschild
line element contains a term proportional to the 
background curvature scalar $R= 6 k$. Thus, it contains a term of order 
$\epsilon^0 \, \delta^2$. The deformed Schwarzschild solution, on the
other hand, contains terms that are of order $\epsilon$ and higher.
Its line element has been worked out in 
\cite{Haack:2008cp,Bhattacharyya:2008mz} at order $\delta^2$, and
there are only two perturbations that are also of order $\epsilon$, namely
the shear tensor $\sigma_{\mu \nu}$ and the perturbation proportional
to $u_{\mu} \, R_{\nu 
\lambda} \, u^{\lambda}$.  The latter contains the term
$u_t \, R_{ij} \, u^j$, which is
of order $\epsilon \, \delta^2$.  At order $\delta^3$, new perturbations
will have to be added to the line element.  Out of these, only perturbations
that are 
proportional to the shear tensor
$\sigma_{\mu \nu}$ can contribute to
$\eta$. 
At order  $\epsilon \, \delta^3$ 
there is only one such
term, namely 
$R \, \sigma_{\mu \nu}$, which for constant $R$ can be absorbed into the
term proportional to $\sigma_{\mu \nu}$ at order $\delta$.
Thus, up to order $\epsilon \, \delta^3$, we 
may restrict the metric perturbations to those involving 
$\sigma_{\mu \nu}$
and to one
particular perturbation of order $\delta^2$ 
associated with the background
curvature, namely $u_{\mu} \, R_{\nu 
\lambda} \, u^{\lambda}$.

Now let us discuss deformed charged black hole solutions. In this
case there are new perturbations present at each order in $\delta$.
For the case of the electrically charged Maxwell black hole, for instance, 
they were computed up to order $\delta^2$ in
\cite{Erdmenger:2008rm,Banerjee:2008th}.
Rather than taking all of these new terms into account, we will follow
the same strategy as in the Schwarzschild case. Namely, we start with 
the deformed solution at order $\delta$ and we add one
particular perturbation of order $\delta^2$ to its line element, 
namely the one  
proportional to $u_{\mu} \, R_{\nu 
\lambda} \, u^{\lambda}$. 

Now that we have clarified the ingredients we need, we make a solution ansatz 
using these
and we 
solve the associated equations of motion up to first order in $\epsilon$.
We do not truncate the equations of motion.
The solution we thus construct at order $\epsilon$ is an exact solution.
It is determined 
in terms of a specific velocity field that is slowly
varying in a certain coordinate range.  
Computing the associated boundary
energy-momentum tensor, we find a correction to $\eta/s$
proportional to the background curvature $k$.
The addition of further deformations
to the line element will, presumably, result in a modified solution that contributes
additional terms to $\eta/s$. 
These new contributions should, however, be
qualitatively different from the one we compute here.

The ratio $\eta/s$ 
should not receive corrections in $\epsilon$, since
that would make it depend on the 
amplitude $\epsilon$ of the velocity field.
Indeed, using the results of \cite{Bhattacharyya:2008mz}, 
we have checked that for the Schwarzschild black hole, 
the second order metric perturbations that are of order 
$\epsilon^2 \, \delta^2$ do not contribute to $\eta$.

The solutions we construct at order $\epsilon$ are based on the specific
velocity field 
\begin{equation}
u^{\mu} = (1, 0,
\epsilon \, \beta^{\phi}(\theta),
\epsilon \, \beta^{\psi}(\theta)) \;.
\label{eq:velo-field-weylz}
\end{equation}
This velocity field has the special feature that
the Weyl connection ${\cal A}_{\mu} = u^{\nu} \, \nabla_{\nu} u_{\mu} 
- \frac13 (\nabla_{\nu} u^{\nu}) \, u_{\mu}$
introduced in \cite{Loganayagam:2008is} vanishes at order $\epsilon$
(here the covariant derivative $\nabla_{\mu}$ is computed using the boundary
metric $g_{\mu \nu}$). In addition, we demand 
that the mass and
the charges of the black hole solution are kept constant at
order $\epsilon \, \delta^2$.

In the following, 
we will first discuss the case of the 
deformed Schwarzschild black hole and then
turn to deformed charged black holes in the STU-model
of $N=2$ gauged supergravity.

\subsection{Deformed Schwarzschild black hole solution}

The construction of a black hole solution dual to a conformal fluid
starts from a stationary black hole solution in Eddington--Finkelstein
coordinates, which then gets deformed by a slowly varying velocity field
\cite{Bhattacharyya:2008jc}.  Let us consider 
the static Schwarzschild solution in Eddington--Finkelstein coordinates
which, according to (\ref{eq:line-schout}), is given by
\begin{equation}
ds^2 = 
- 2\, u_{\mu} \, dx^{\mu} \, da + \left[
a^2  \, g_{\mu \nu} + 
u_{(\mu} \, S_{\nu) \lambda} \, u^{\lambda} +  \frac{\mu}{a^2}
\, u_{\mu} \, u_{\nu}
\right]
dx^{\mu} \, 
dx^{\nu} \;,
\label{eq:schwarz-static}
\end{equation}
where $u_{\mu} = (-1,0,0,0)$.
Observe that the term proportional to the Schouten tensor is
of order $\epsilon^0 \, \delta^2$.
The associated function $f$ reads $f= 1 + k/a^2 - \mu/a^4 $.  The event
horizon is at $f(a_h)=0$.  It will be useful to introduce rescaled
variables $\rho = a/a_h$ and $m = \mu/a^4_h$, in terms of which $f$ is
given by
\begin{equation}
f(\rho) = 1 + \frac{k}{a^2_h \, \rho^2} - \frac{m}{\rho^4} \;.
\end{equation}
The event horizon is at $\rho = 1$ and $m$ satisfies $m = 1 + k/a^2_h $.

Now we deform (\ref{eq:schwarz-static}) by taking the 
velocity field to be non-trivial.  The perturbed line element
is then written in terms of Weyl covariant combinations 
\cite{Loganayagam:2008is,Bhattacharyya:2008mz}.
We work at first order in $\epsilon$, and we take the velocity field
to be of the form (\ref{eq:velo-field-weylz}), for which the
Weyl connection vanishes at first order in $\epsilon$.
The vanishing of the latter implies that
the Weyl-covariantized Schouten tensor ${\cal S}_{\mu \nu}$ 
coincides with the ordinary Schouten tensor $S_{\mu \nu}$. 

In general, when deforming the static black hole solution, not
only the velocity field $u^{\mu}$ but also 
the mass $\mu$ becomes a slowly varying 
function of $x^{\mu}$
\cite{Bhattacharyya:2008jc}.  
For the velocity field (\ref{eq:velo-field-weylz}),
inspection of equation (C.1) in \cite{Bhattacharyya:2008mz}
shows that $\mu$ remains constant at order $\epsilon \, \delta^2$
provided that
${\cal D}^{\nu} \sigma_{\nu \mu} =0$.  Here
${\cal D}$
denotes the Weyl covariant derivative introduced in  
\cite{Loganayagam:2008is}, and the
shear tensor $\sigma_{\mu \nu}$ is defined below.
Using this information, we make an ansatz for the line element
that captures effects of
order $\epsilon \, \delta^2$,
and we take $\mu$ to be constant.

We deform 
(\ref{eq:schwarz-static})
by adding a term proportional
to the shear tensor $\sigma_{\mu \nu}$ \cite{Bhattacharyya:2008jc,
Loganayagam:2008is,Bhattacharyya:2008mz},
\begin{equation}
\sigma_{\mu \nu} = \frac12 \left({\cal D}_{\mu} u_{\nu} 
+ {\cal D}_{\nu} u_{\mu} \right) = 
\frac12 \left( P_{\mu \lambda} \, 
\nabla^{\lambda} u_{\nu} + 
 P_{\nu \lambda} \, 
\nabla^{\lambda} u_{\mu} \right) - \frac13 \, P_{\mu \nu} \,
\nabla_{\lambda} u^{\lambda} \;,
\end{equation}
where $P_{\mu \nu} = g_{\mu \nu} + u_{\mu} u_{\nu}$. 
For the velocity field (\ref{eq:velo-field-weylz}) this yields
$\sigma_{\mu \nu} = \frac12 \left({\nabla}_{\mu} u_{\nu} 
+ {\nabla}_{\nu} u_{\mu} \right)$ to first order in $\epsilon$.
Thus we make the following 
ansatz for the
perturbed line element at order $\epsilon$,
\begin{eqnarray}
ds^2 &=& 
- 2\, u_{\mu} \, dx^{\mu} \, da + \left[
a^2  \, g_{\mu \nu} + 
u_{(\mu} \, S_{\nu) \lambda} \, u^{\lambda} +  \frac{\mu}{a^2}
\, u_{\mu} \, u_{\nu}
\right]
dx^{\mu} \, 
dx^{\nu} \nonumber\\
&& + 
2 \, \frac{a^2}{a_h} \, F(a) \,
\sigma_{\mu \nu}  \,
dx^{\mu} \, dx^{\nu} \;.
\label{eq:line-ansatz-max-schwarz}
\end{eqnarray}
Here, $F$ has Weyl weight zero, so that (\ref{eq:line-ansatz-max-schwarz})
is invariant under the rescalings (\ref{eq:scalings}).
Observe that according to the counting described above,
$\sigma_{\mu \nu}$ is of order $\epsilon \, \delta$, while 
$u_{(\mu} \, S_{\nu) \lambda} \, u^{\lambda}$ contains the deformation term 
$u_{(\mu} \, R_{\nu) \lambda} \, u^{\lambda}$
which is of order $\epsilon \,\delta^2$.

Imposing the condition ${\cal D}^{\nu} \sigma_{\nu \mu} =0$ we find
the following expression for the velocity field,
\begin{eqnarray}
\beta^{\phi}(\theta)&=& \omega_1 + c_1
\left(-\frac14\log[\cos\,\theta]
+\frac14\log[\sin\,\theta]+\frac{1}{8\, \cos^2\,\theta}\right) \;,
\nonumber\\
\beta^{\psi}(\theta)&=& \omega_2
+c_2 \left(-\frac14\log[\cos\,\theta]
+\frac14\log[\sin\,\theta]-\frac{1}{8 \, \sin^2\,\theta}\right) \;,
\label{eq:velocity-field}
\end{eqnarray}
with constants $\omega_1, \omega_2, c_1, c_2$. 
Observe that in obtaining (\ref{eq:velocity-field}) we have not resorted
to any approximation, 
i.e. at order $\epsilon$ (\ref{eq:velocity-field})
solves  ${\cal D}^{\nu} \sigma_{\nu \mu} =0$ exactly.
The small amplitude approximation, however, breaks down at 
$\theta = 0, \pi/2$, where the norm of the velocity field
diverges.  Therefore, 
we have to restrict the range of $\theta$ to be consistent with
the small amplitude expansion. This may be achieved by restricting
$\theta$ to be in the range 
$\lambda < \theta < \pi/2 -  \lambda$ with $\epsilon << \lambda^2 $.

In case that both the $c_i$
($i=1,2$) vanish, 
(\ref{eq:line-ansatz-max-schwarz}) describes an uncharged stationary
black hole solution (at order $\epsilon$)
with $\sigma_{\mu \nu} =0$.
In the following, we will be interested in non-stationary solutions,
and hence we take at least one of the $c_i$ to be non-vanishing.
Using (\ref{eq:velocity-field}), and 
inserting the ansatz (\ref{eq:line-ansatz-max-schwarz}) into the 
Einstein equations of motion, we find that they are satisfied to
first order in $\epsilon$
provided that 
$F$ satisfies the differential equation 
\begin{equation}
\frac{d}{d \rho} \left( \rho^5 \, f(\rho) \, \frac{d}{d \rho} F(\rho) 
\right) = 
- \left( 3 \rho^2 + \frac{k}{a^2_h} \right) \;.
\label{eq:diff2Fschw}
\end{equation}
When solving the Einstein equations, we do not
resort to any truncation. 
Thus, 
(\ref{eq:velocity-field}) and
(\ref{eq:diff2Fschw}) yield an exact solution to 
the Einstein equations at first order in $\epsilon$.

Integrating (\ref{eq:diff2Fschw}) once gives
\begin{equation}
\rho^5 \, f(\rho) \, \frac{d}{d \rho} F = 
- \left( \rho^3 + \frac{k}{a^2_h} \, \rho - \zeta \right) \;,
\label{eq:diffFschw}
\end{equation}
where the integration constant $\zeta$ is set to the value 
$\zeta = 1+ k/a^2_h$
so as to account for the vanishing of $f(\rho)$ at the horizon 
$\rho = 1$.  Note that (\ref{eq:diffFschw}) can be written as
\begin{equation}
\frac{d}{d \rho} F = - 
\frac{\left(\rho^2 + \rho + \zeta \right)}{\rho 
(\rho + 1) (\rho^2 + \zeta 
)} \;.
\label{eq:Fdiff-schwarz}
\end{equation}
Integrating (\ref{eq:Fdiff-schwarz}) once results in 
\begin{equation}
F(\rho) = \int_{\rho}^{\infty} du \, 
\frac{ \left(u^2 + u + \zeta \right)}{u (u + 1) (u^2 + \zeta ) } \;,
\label{eq:result-F}
\end{equation}
which is well-behaved as long as $\rho > 0$.  
In the limit of large $\rho$ this yields
\begin{equation}
\frac{F(\rho)}{a_h} = \frac{1}{a} - \frac{\eta}{4 \, a^4}  \;,
\end{equation}
where $\eta = \zeta \, a^3_h = a^3_h + k \, a_h$.

Next we consider the 
fluid on a three-sphere dual to 
(\ref{eq:line-ansatz-max-schwarz}).
Its
energy-momentum tensor $T_{\mu \nu}$ can be computed 
using standard techniques 
\cite{Balasubramanian:1999re,Emparan:1999pm,Batrachenko:2004fd}, see appendix 
\ref{tmnSTU}.
We obtain
\begin{eqnarray}\label{tmn-schwarz}
16\pi G_5\, \langle T_{\mu\nu} \rangle
&=& \frac{1}{4}\left(R_{\alpha\beta} 
R^{\alpha\;\;\;\beta}_{\,\;\,\mu\;\;\;\nu} -\frac{R^2 }{12}g_{\mu\nu}\right) 
\\ \nonumber 
&&+ \mu\left(g_{\mu\nu}+4\,u_\mu\,u_\nu\right)-
2\,\eta \,\sigma_{\mu\nu}\;.
\end{eqnarray}
The terms in the first line of this expression denote
the contribution to the energy-momentum 
tensor of global $AdS_5$ \cite{Balasubramanian:1999re,Bhattacharyya:2008ji}, 
while
the terms proportional to $\mu$ denote the 
perfect fluid contribution ($\mu$ is related to the pressure 
$p = M/(3 \, vol (S^3))$ by
$\mu = 16 \pi \, G_5 \, p$).
The last term is proportional to the shear tensor.
In units where $L = 16 \pi G_5 =1$
the entropy density $s$ of the fluid 
on a unit three-sphere
is $s = {\cal S}/{vol (S^3)} = 4 \pi \, a^3_h$, 
so that the ratio $\eta/s$ reads
\begin{equation}
\frac{\eta}{s} = \frac{1}{4 \pi} \left( 1 + \frac{k}{a^2_h} \right) \;.
\end{equation}

\subsection{Deformed Maxwell black hole solution}

Next, let us consider the Maxwell black hole in the context of the STU-model.
To this end, we
set $X^1 = X^2 = X^3 =1$ as well as $A^1 = A^2 = A^3 = 2 A/\sqrt{3}$.
Then, from (\ref{eq:line-schout}), we obtain the 
following line element for the static Maxwell black hole,
\begin{equation}
ds^2 = 
- 2\, u_{\mu} \, dx^{\mu} \, da + \left[
a^2  \, g_{\mu \nu} + 
u_{(\mu} \, S_{\nu) \lambda} \, u^{\lambda} +  \left(\frac{w_5 \, M}{a^2}
- \frac{Q^2}{a^4} \right)
\, u_{\mu} \, u_{\nu}
\right]
dx^{\mu} \, 
dx^{\nu} \;.
\label{eq:line-max-static}
\end{equation}
The Maxwell gauge potential reads
\begin{equation}
A_{\mu} = - \frac{\sqrt{3}}{2} \,
\frac{Q}{a^2} \, u_{\mu} \;\;\;,\;\;\; A_a = 0 \;,
\label{eq:gaugepot-max}
\end{equation}
where $u_{\mu} = (-1,0,0,0)$.
The function $f$ in (\ref{eq:function-f}) reads
$f(a) = 1 + k/a^2 - w_5 \, M/a^4 + Q^2/a^6 $.
The location $a_h$ of the outer event horizon is given by the largest positive 
root of $f(a)$.  In terms of the
rescaled variables $\rho = a/a_h$,
 $m = w_5 \, M/a^4_h$ and ${\cal Q} = 
Q/a^3_h$, the function $f$ is given by
\begin{equation}
f(\rho) = 1 + \frac{k}{a^2_h \, \rho^2} - \frac{m}{\rho^4} +
\frac{{\cal Q}^2}{\rho^6} \;.
\label{eq:frho-Max}
\end{equation}
The outer event horizon is at $\rho =1$ and $m$ satisfies
$m = 1 + k/a^2_h + {\cal Q}^2$.

Now we deform the static Maxwell solution by taking the 
velocity field to be of the form (\ref{eq:velo-field-weylz}) with
$\beta^{\phi}$ and $\beta^{\psi}$ 
given by (\ref{eq:velocity-field}).
We work at first order in $\epsilon$, as before.
The results of \cite{Erdmenger:2008rm,Banerjee:2008th} 
show that at order $\epsilon \,
\delta^2$, the electric charge $Q$ can be kept constant 
when $M$ is constant.  In the following, we take both $M$ and
$Q$ to be constant.

We construct a solution to the combined Einstein-Maxwell equations of
motion as follows. We take the gauge potential
to be of the form (\ref{eq:gaugepot-max}) with the velocity field
given by (\ref{eq:velocity-field}). Inserting this ansatz into the equations
of motion, we find that we can solve the combined system exactly
at first order in $\epsilon$ with the following 
line element,
\begin{eqnarray}
\label{line-maxdef-1}
ds^2&=& -2\,u_\mu\,dx^\mu\,da+ \left[a^2\,g_{\mu\nu} + 
\,u_{(\mu }\,S_{\nu)\lambda}\,u^\lambda 
+\left(\frac{w_5 \, M}{a^2} - \frac{Q^2}{a^4}\right)u_\mu\,u_\nu
\right]\,dx^\mu\,dx^\nu  
\nonumber\\
& &+ \left[
2 \sqrt{3} \, \kappa \,
\frac{Q}{a^2}\, u_{(\mu}\,l_{\nu)} 
+  
2 \, \frac{a^2}{a_h} \, F(a) \,
\sigma_{\mu \nu} 
\right]\,dx^\mu\,dx^\nu 
\nonumber\\
&& 
+
4 \sqrt{3} \, \kappa \,
\frac{Q}{a^4\,f(a)}\,l_\mu\,dx^\mu\,da \;,
\end{eqnarray}
where we recall that $ u_{(\mu}\,l_{\nu)} = u_{\mu} \, l_{\nu}
+ u_{\nu} \, l_{\mu}$, and where 
\cite{Bhattacharyya:2008ji,Erdmenger:2008rm,Banerjee:2008th} 
\begin{equation}
l_{\mu} = \frac12 \, \epsilon_{\mu \nu \lambda \sigma} \, u^{\nu} \, {\cal D}^{\lambda}
u^{\sigma} = \frac12 \, \epsilon_{\mu \nu \lambda \sigma} \, u^{\nu} \, \nabla^{\lambda}
u^{\sigma}  \;,
\label{def-l}
\end{equation}
with $\epsilon_{\mu \nu \lambda \sigma} = e_{\mu}{}^a \, 
e_{\nu}{}^b \, e_{\lambda}{}^c \, e_{\sigma}{}^d \, \epsilon_{a b c d}$.
Observe that $l_{\mu}$ and $F(a)$ have Weyl-weight zero, and that the 
associated terms in (\ref{line-maxdef-1}) are of order $\epsilon \, \delta$,
while 
$u_{(\mu} \, S_{\nu) \lambda} \, u^{\lambda}$ contains the deformation term 
$u_{(\mu} \, R_{\nu) \lambda} \, u^{\lambda}$
which is of order $\epsilon \,\delta^2$. The line element (\ref{line-maxdef-1})
is invariant under the rescalings (\ref{eq:scalings}) and (\ref{eq:scalings2}).

The quantity $F$ now satisfies the differential equation 
\begin{equation}
\frac{d}{d \rho} \left( \rho^5 \, f(\rho) \, \frac{d}{d \rho} F(\rho) 
\right) = 
- \left( 3 \rho^2 + \frac{k}{a^2_h} \right) \;,
\label{eq:diff2F}
\end{equation}
with $f(\rho)$ given by (\ref{eq:frho-Max}).
Integrating (\ref{eq:diff2F}) once gives
\begin{equation}
\rho^5 \, f(\rho) \, \frac{d}{d \rho} F = 
- \left( \rho^3 + \frac{k}{a^2_h} \, \rho - \zeta \right) \;,
\label{eq:diffF}
\end{equation}
where the integration constant $\zeta$ is set to the value 
$\zeta = 1+ k/a^2_h$
so as to account for the vanishing of $f(\rho)$ at the outer horizon 
$\rho = 1$.  Note that (\ref{eq:diffF}) can be written as
\begin{equation}
\frac{d}{d \rho} F = - 
\frac{\rho \left(\rho^2 + \rho + \zeta \right)}{(\rho + 1) (\rho^4 + \zeta \,
\rho^2 - {\cal Q}^2)} \;.
\label{eq:Fdiff}
\end{equation}
Integrating (\ref{eq:Fdiff}) once results in 
\begin{equation}
F(\rho) = \int_{\rho}^{\infty} du \, 
\frac{u \left(u^2 + u + \zeta \right)}{(u + 1) (u^4 + \zeta \,
u^2 - {\cal Q}^2)} \;.
\end{equation}
Here $\rho$ should be taken to be larger than the largest positive root of 
$u^4 + \zeta \,u^2 - {\cal Q}^2$ to avoid a singularity in $F(\rho)$.
In the limit of large $\rho$ this yields
\begin{equation}
 \frac{F(\rho)}{a_h} = \frac{1}{a} - \frac{\eta}{4 \, a^4}  \;,
\label{eq:Fasym-mxw1}
\end{equation}
where $\eta = \zeta \, a^3_h = a^3_h + k \, a_h$.

The line element (\ref{line-maxdef-1}) is not in the customary gauge
$g_{a \mu} = - u_{\mu}$ \cite{Bhattacharyya:2008mz}. It can be brought
into this gauge by the following coordinate transformation at order $\epsilon$,
\begin{equation}
dx^{\mu} \rightarrow dx^{\mu} - h(a) \, l^{\mu} \, da - 
\left( \int^a h(b) \, db \right) \, d l^{\mu} \;,
\label{eq:min-gauge}
\end{equation}
where $h(a) =  2 \sqrt{3} \, \kappa \, Q/(a^6 \, f(a) )$. Here the term
proportional to $l^{\mu}$ is of order $\epsilon \, \delta$, while the 
term proportional to $d l^{\mu}$ is of order $\epsilon \, 
\delta^2$.
The resulting line element is then regular at the outer horizon $f(a_h) =0$ of
the undeformed static black hole solution.

In the stationary case, the velocity field has the form
(\ref{eq:velocity-field}) with $c_i =0$. Due to the 
curvature $k$ of the background, $l^{\mu}$ is non-vanishing but constant and 
given by 
$l^{\mu} = \sqrt{k} \, (0,0,-\omega_2 ,-\omega_1)$.  
Then
the second term in (\ref{eq:min-gauge}) vanishes, and the
line element takes the form 
\begin{eqnarray}
ds^2 &=& 
- 2\, u_{\mu} \, dx^{\mu} \, da + \left[
a^2  \, g_{\mu \nu} + 
u_{(\mu} \, S_{\nu) \lambda} \, u^{\lambda} +  \left(\frac{w_5 \, M}{a^2}
- \frac{Q^2}{a^4} \right)
\, u_{\mu} \, u_{\nu} \right. \nonumber\\
&& \hskip 3cm 
\left. + 2 \sqrt{3} \, \kappa \,
\frac{Q}{a^2}\,u_{(\mu}\,l_{\nu)}
\right]\,dx^\mu\,dx^\nu 
\end{eqnarray}
in the gauge $g_{a \mu} = - u_{\mu}$.
It is straightforward to relate this line element to the usual
one \cite{Chong:2005hr} written in Boyer--Lindquist type coordinates, 
to linear order in $\omega_1$ and $\omega_2$,
see appendix \ref{bl-ef}.

Next we compute the associated
boundary energy-momentum tensor $T_{\mu \nu}$ of the fluid,
see appendix \ref{tmnSTU}. 
We obtain 
\begin{eqnarray}
\label{eq:tmn-max1}
16\pi G_5\, \langle T_{\mu\nu} \rangle
&=& \frac{1}{4}\left(R_{\alpha\beta} 
R^{\alpha\;\;\;\beta}_{\,\;\,\mu\;\;\;\nu} -\frac{R^2 }{12}g_{\mu\nu}\right) 
\\ \nonumber 
&&+ w_5 \, M \left(g_{\mu\nu}+4\,u_\mu\,u_\nu\right)
+ 8 \sqrt{3} \, \kappa \, Q \, u_{(\mu} \, l_{\nu)} 
-2
\,\eta \,\sigma_{\mu\nu}
\;.
\end{eqnarray}
In units where $L = 16 \pi G_5 =1$ (using that
the entropy density $s$ of the fluid 
on a unit three-sphere
is $s = {\cal S}/{vol (S^3)} = 4 \pi \, a^3_h$), 
the ratio $\eta/s$ reads
\begin{equation}
\frac{\eta}{s} = \frac{1}{4 \pi} \left( 1 + \frac{k}{a^2_h} \right) \;,
\label{eq:shear-mxw1}
\end{equation}
as in the Schwarzschild case.
We note that the correction to $\eta/s = 1/(4 \pi)$ is determined
by the coefficient of the $u \, R \, u$-term in the line 
element \eqref{line-maxdef-1}.

In the stationary case, where $\sigma_{\mu \nu} =0$, 
$T_{\mu \nu}$ takes the form
given in \cite{minwallastrings}.  It contains additional 
non-dissipative terms proportional
to $l_{\mu}$ associated with the rotation of the fluid in a
background of constant curvature $k$.

In \cite{Erdmenger:2008rm,Banerjee:2008th} the authors constructed
charged black brane solutions up to order $\delta^2$.  
At order $\delta$, their solution is based on the gauge field
\begin{equation}
\label{gaugefield2}
A_{\mu} =
- \frac{\sqrt{3}\,Q}{2\,a^2}\, \left(
u_\mu - 2 \sqrt{3}\,\kappa\, \frac{Q}{w_5 \, M} \, l_\mu \right) 
\;\;\;,\;\;\; A_a= 0 \;.
\end{equation}
For the sake of comparison, let us construct a black hole solution
based on (\ref{gaugefield2})
with the velocity field
given by (\ref{eq:velocity-field}). Inserting this ansatz into the equations
of motion, we find that we can solve them exactly
at first order
in $\epsilon$ 
with the following 
line element,
\begin{eqnarray}
ds^2&=&-2\,u_\mu\,dx^\mu\,da+ \left[a^2\,g_{\mu\nu} + 
\,u_{(\mu} \,S_{\nu )\lambda}\,u^\lambda 
+ \left(\frac{w_5 \, M}{a^2} - \frac{Q^2}{a^4}\right)u_\mu\,u_\nu
 \right]\,dx^\mu\,dx^\nu \nonumber\\
& &+\left[
- \frac{6\,\kappa^2 Q^2}{w_5 \, M\,a^2} \,
u_{(\mu}\,R_{\nu) \lambda}\,u^\lambda 
+ \frac{2\sqrt{3}\,\kappa\,Q^3}{w_5 \, M\,a^4}\,
u_{(\mu}\,l_{\nu)} 
+2\,
\frac{a^2}{a_h}\,F(a)\,\sigma_{\mu\nu}
\right]\,dx^\mu\,dx^\nu \nonumber\\ 
& &+\left[\frac{4\sqrt{3}\,\kappa\,Q^3}{w_5 \, M\,a^6\,f}\,l_\mu -
  \frac{12\,\kappa^2\,Q^2}{w_5 \, M\,a^4
    f}\,R_{\mu\lambda}\,u^\lambda\right]\,dx^\mu\,da \;,
\label{line-maxdef-2}
\end{eqnarray}
with $l_{\mu}$ defined as in (\ref{def-l}).  The quantity $F$ satisfies 
the differential equation (\ref{eq:diff2F}).
The line element (\ref{line-maxdef-2}) is invariant under the rescalings
(\ref{eq:scalings}) and (\ref{eq:scalings2}).
It is again not in the customary gauge
$g_{a \mu} = - u_{\mu}$ \cite{Bhattacharyya:2008mz}. It can be brought
into this gauge by the coordinate transformation (\ref{eq:min-gauge})
at order $\epsilon$. The resulting line element 
is then regular at the outer horizon $f(a_h) =0$ of
the undeformed static black hole solution.

One may ask whether the two line elements (\ref{line-maxdef-1})
and (\ref{line-maxdef-2}) can be transformed into each other.
The associated gauge
fields are related by the shift 
$u^{\mu} \rightarrow 
u^{\mu} - 2 \sqrt{3} \, \kappa \,  \frac{Q}{w_5 \, M} \,
l^{\mu} $.  Applying this shift to the line element
(\ref{line-maxdef-1}) induces terms that are of order $\epsilon \, \delta^3$.
The resulting line element thus has terms of different order in $\delta$
than the line element (\ref{line-maxdef-2}).  Matching of these two line
elements is thus only expected to occur when the 
full set of $\epsilon \, \delta^3$-terms is taken into account.
However, in the stationary case
($c_i =0$), the solution (\ref{line-maxdef-1}) 
is mapped into 
 (\ref{line-maxdef-2}) at order $\epsilon$
by the shift of $u^{\mu}$ described above, under
which 
$l_i \rightarrow l_i 
- \sqrt{3} \, \kappa \, \frac{Q}{w_5 \,M } \, R_{ij} \, u^j$.
The two line elements are then identical in the gauge $g_{a \mu} = - u_{\mu}$,
as expected.

Let us now compare the line element (\ref{line-maxdef-2}) with 
the one obtained in \cite{Erdmenger:2008rm,Banerjee:2008th}. Since
the gauge field (\ref{gaugefield2}) is at most of order $\epsilon \, \delta$,
the comparison is only meaningful up to this order. Since
the terms in (\ref{line-maxdef-2}) proportional to $R_{\mu \nu}$ are of
order $\epsilon \, \delta^2$ they should be dropped in the comparison.
Then, by going into the 
gauge $g_{a \mu} = - u_{\mu}$ 
via the coordinate transformation (\ref{eq:min-gauge}) (and dropping the
term proportional to $d l^{\mu}$ which is also 
of order $\epsilon \, \delta^2$) 
we find that the line element (\ref{line-maxdef-2}) goes over into the
one obtained in \cite{Erdmenger:2008rm,Banerjee:2008th}.

Computing the associated
boundary energy-momentum tensor $T_{\mu \nu}$ we
obtain
\begin{eqnarray}
\label{eq:tmn-max2}
16\pi G_5\, \langle T_{\mu\nu} \rangle
&=& \frac{1}{4}\left(R_{\alpha\beta} 
R^{\alpha\;\;\;\beta}_{\,\;\,\mu\;\;\;\nu} -\frac{R^2 }{12}g_{\mu\nu}\right) 
\\ \nonumber 
&&+ w_5 \, M \,  \left(g_{\mu\nu}+4\,u_\mu\,u_\nu\right)
-\frac{24\,\kappa^2\,Q^2}{w_5 \, M} \, u_{(\mu} \, R_{\nu )\lambda}\, 
u^\lambda 
-2
\,\eta \,\sigma_{\mu\nu}\;,
\end{eqnarray}
with $\eta/s$ given by (\ref{eq:shear-mxw1}).
It contains non-dissipative 
terms proportional to the background curvature tensor 
$R_{\mu \nu}$.
In the stationary case,  the 
boundary energy-momentum tensor (\ref{eq:tmn-max1}) matches (\ref{eq:tmn-max2})
under the constant shift
$u^{\mu} \rightarrow u^{\mu} - 2 \sqrt{3} \, \kappa \,  \frac{Q}{w_5 \, M} \,
l^{\mu}$ discussed above.

\subsection{Deformed black hole solutions supported by scalar fields}
\label{BlackHoleScalar}

Next, we consider black hole solutions in the STU-model that
are supported by non-trivial scalar fields, and that carry either
one or two non-vanishing charges. In the two-charge case, we take
the charges to be equal, for simplicity.  
We deform the static solutions in the manner described above.
We find that
the scalar fields need not be deformed at order $\epsilon$.

\subsubsection{Two equal charges}
\label{STU2charges}

We begin by first considering the case of two equal charges.
The line element of the static solution is given by
(\ref{eq:line-boost1})
and the gauge potentials and scalar fields are 
\begin{eqnarray}\label{staticfields}
 A_\mu^1 &=& A_\mu^2 = -\frac{Q}{a^2}\,H^{-\frac13}\, u_\mu \,\,,\quad A_\mu^3 = 0
\,\,,\quad A_a^i = 0 \,\,,\quad i=1,2,3 \quad , \nonumber \\ 
X^1 &=& X^2=H^{-\frac13} \,\,,\quad X^3=H^\frac23 \,,
\end{eqnarray}
where $u_\mu=(\,-1,\,0,\,0,\,0)$. We refer to the end of section
\ref{elec-bhs-static} for a definition of
 the various quantities involved.  The function $f(a)$ appearing in (\ref{eq:line-boost1}), when 
expressed in terms of the rescaled coordinates $\rho=a/a_h$, reads
\begin{eqnarray}
f(\rho) = 1+{\rm e}^{-4U}\,\frac{k}{a_h^2 \,\rho^2}-
{\rm e}^{-2U}\,\frac{m}{\rho^4} \;\;\;,\;\;\;
m=\frac{\mu}{a_h^4}=\left(1+\frac{k}{a_h^2}\,{\rm e}^{-4U(a_h)}\right)\,{\rm e}^{2U(a_h)} \;.
\end{eqnarray}
The outer horizon is at $\rho=1$.

We perturb this static solution by again taking the velocity field 
to have the form (\ref{eq:velo-field-weylz}) and 
(\ref{eq:velocity-field}). This results in a modification of 
the line element, and it also induces a non-vanishing $A^3$.
We find that at first order in $\epsilon$ (but no approximation
otherwise)
the combined system of equations of motion
is solved by
\begin{eqnarray}\label{STUblackPert}
ds^2&=&-a^2\,f(a)\,u_\mu\,u_\nu\,dx^\mu\,dx^\nu - \frac{6}{W(a)}\,u_\mu\,dx^\mu\,da 
+ a^2\,P_{\mu\nu}\,dx^\mu\,dx^\nu \nonumber \\  
& &+ \frac12\,H^{-\frac13}\,u_{(\mu}\,R_{\nu)\lambda}\,u^\lambda 
\,dx^\mu\,dx^\nu 
+ 2\,\frac{a^2}{a_h}\,F(a)\,\sigma_{\mu\nu}\,dx^\mu\,dx^\nu\,, \nonumber \\
A_\mu^1 &=& A_\mu^2 = -\frac{Q}{a^2}\,H^{-\frac13}\, u_\mu \,\,,\quad 
A_\mu^3 = -\frac{q}{a^2}\,H^{\frac23}\,l_\mu \,\,,\quad A_a^i = 0 \,\,,\quad i=1,2,3 \quad ,
\end{eqnarray}
with the scalar fields given as in \eqref{staticfields}. Here $l_\mu$ and 
the velocity field are again given by \eqref{def-l} and (\ref{eq:velocity-field}), respectively. 
The stationary limit of this solution can be easily related to the solution found in \cite{Chong:2005da} 
written in Boyer--Lindquist type coordinates, to linear order in rotation parameters (see appendix \ref{STUblackhole}).

The quantity $F$ now satisfies the differential equation
\begin{equation}\label{diffF} 
\frac13\,\frac{d}{d\rho}\,\left(\rho^5\,W(\rho)\,f(\rho)\,\frac{d}{d\rho}F(\rho)\right) 
= -\left(3\rho^2 + \frac{k}{a_h^2}\,{\rm e}^{-U}\,(1-U'\rho)\right) \,,
\end{equation}
where $U'=d U/d\rho$, with ${\rm e}^{3U} = H$. 
We note the appearance of the superpotential $W(a)$ on the 
left hand side,
which was constant ($W(a) = 3$) in both the Schwarzschild and
the Maxwell case. The right hand side of \eqref{diffF}
can be easily integrated by noting that the second term
is a total derivative, 
\begin{equation}
 {\rm e}^{-U}\,(1-U'\rho)\,d\rho = d\left(\rho\,{\rm e}^{-U}\right) \,.
\end{equation}
Thus, 
integrating \eqref{diffF} once gives
\begin{eqnarray}\label{diffF2}
\frac13\,\rho^5\,W(\rho)\,f(\rho)\,\frac{d}{d\rho}F(\rho) = 
-\left(\rho^3 + \frac{k}{a_h^2}\,{\rm e}^{-U}\rho - \zeta \right) \,.
\end{eqnarray}
The integration constant $\zeta$ is set to the value 
$\zeta=1+ \left( k\,{\rm e}^{-U(a_h)}\right)/a^2_h$ to allow for the vanishing of \eqref{diffF} 
at the outer horizon $\rho=1$, where $f=0$. Then, integrating
\eqref{diffF2} once results in 
\begin{eqnarray}
F(\rho) =\int_{\rho}^{\infty} 
\frac{du}{W(u)}\,\frac{3\left( u^3 + u\,(\zeta - 1)\,{\rm e}^{U(a_h)}
\,{\rm e}^{-U} - \zeta \right)}
{u^5 + u^3\,(\zeta -1)\,{\rm e}^{U(a_h)}\,{\rm e}^{-4U} - u\,\left({\rm e}^{2U(a_h)}+(\zeta - 1)\,{\rm e}^{-U(a_h)}\right)\,{\rm e}^{-2U}} .
\end{eqnarray}
For large $\rho$ we have ${\rm e}^{3U} = H \approx
1 + q \, /(a_h^2 \,\rho^2)$, and hence we obtain
\begin{eqnarray}\label{FSTU}
 \frac{F(\rho)}{a_h} = \frac{1}{a} -\frac{\eta}{4\,a^4} \,,
\end{eqnarray}
 where $\eta=\zeta\,a_h^3=a_h^3+k\,{\rm e}^{-U(a_h)}\,a_h$. 

Computing the associated boundary energy-momentum tensor 
we obtain (see appendix \ref{tmnSTU})
\begin{eqnarray}
\label{eq:tmn-STU}
16\pi G_5\, \langle T_{\mu\nu} \rangle
&=& \frac{1}{4}\left(R_{\alpha\beta} 
R^{\alpha\;\;\;\beta}_{\,\;\,\mu\;\;\;\nu} -\frac{R^2 }{12}g_{\mu\nu}\right) \\ \nonumber 
&&+ w_5\,M\,\left(g_{\mu\nu}+4\,u_\mu\,u_\nu\right)-\frac{2\,q}{3}\,u_{(\mu}\,R_{\nu)\lambda}\,u^\lambda -2\,\eta \,\sigma_{\mu\nu} \,.
\end{eqnarray}
It contains a non-dissipative term proportional to the
background curvature tensor $R_{\mu \nu}$.
In units where $L = 16 \pi G_5 =1$, 
the ratio $\eta/s$ reads
\begin{equation}
\frac{\eta}{s} = \frac{1}{4 \pi} \left( 1 + \frac{k \,
{\rm e}^{-U(a_h)}}{a^2_h} \right) \;.
\label{eq:shear-stu-2}
\end{equation}
We note that the correction to $\eta/s = 1/(4 \pi)$ is determined
by the coefficient of the $u \, R \, u$-term in the line 
element \eqref{STUblackPert}.

\subsubsection{One charge}
\label{STU1charge}

Next we consider the case of one non-vanishing charge.
Proceeding as before, i.e. taking the velocity field to be 
given by (\ref{eq:velocity-field}), 
we find that at first order in $\epsilon$
(but no approximation otherwise)
the perturbed solution
to the combined system of equations of motion is given by
\begin{eqnarray}\label{STUblackPert2}
ds^2&=&-a^2\,f(a)\,u_\mu\,u_\nu\,dx^\mu\,dx^\nu - \frac{6}{W(a)}\,u_\mu\,dx^\mu\,da 
+ a^2\,P_{\mu\nu}\,dx^\mu\,dx^\nu \nonumber \\  
& &+ \frac12\,H^{\frac13}\,u_{(\mu}\,R_{\nu)\lambda}\,u^\lambda 
\,dx^\mu\,dx^\nu 
+ 2\,\frac{a^2}{a_h}\,F(a)\,\sigma_{\mu\nu}\,dx^\mu\,dx^\nu\,, \nonumber \\
A_\mu^1 &=&-\frac{Q}{a^2}\,H^{-\frac23}\, u_\mu \,\,,\quad  A_\mu^2=A_\mu^3 =0 
\,\,,\quad A_a^i = 0 \,\,,\quad i=1,2,3 \quad , \nonumber \\
X^1&=&H^{-\frac23} \,,\quad X^2=X^3=H^\frac13 \,.
\end{eqnarray}
The stationary limit of this solution can be related to the solution found in \cite{Cvetic:2004ny,Chong:2006zx}, 
to linear order in rotation parameters (see appendix \ref{genSTU}). 
 The quantity $F$ satisfies the differential equation
\begin{equation}\label{diffFSTU2} 
\frac13\,\frac{d}{d\rho}\,\left(\rho^5\,W(\rho)\,f(\rho)\,\frac{d}{d\rho}F(\rho)\right) 
= -\left(3\rho^2 + \frac{k}{a_h^2}\,{\rm e}^{2U}\,(1+2\,U'\rho)\right) \,,
\end{equation}
where $U'=dU/d\rho$, with 
${\rm e}^{6U} = H$. 
The right hand side of \eqref{diffFSTU2}
can be easily integrated by noting that the second term
is a total derivative, 
\begin{equation}
 {\rm e}^{2U}\,(1+2\,U'\rho)\,d\rho = d\left(\rho\,
 {\rm e}^{2U}\right) \,.
\end{equation}
Integrating \eqref{diffFSTU2} once gives
\begin{eqnarray}\label{diffF2STU2}
\frac13\,\rho^5\,W(\rho)\,f(\rho)\,\frac{d}{d\rho}F(\rho) = 
-\left(\rho^3 + \frac{k}{a_h^2}\,{\rm e}^{2U}\rho - \zeta \right) \,.
\end{eqnarray}
The integration constant $\zeta$ is set to the value 
$\zeta=1+\left(k \,{\rm e}^{2U(a_h)}\right)/a^2_h$ to allow for the vanishing of \eqref{diffFSTU2} 
at the outer horizon $\rho=1$, since $f=0$ there. Then, 
integrating \eqref{diffF2STU2} once results in 
\begin{eqnarray}
F(\rho) =\int_{\rho}^{\infty} \frac{du }{W(u)}\,\frac{
3 \left(u^3 + u\,(\zeta - 1)\,{\rm e}^{-2U(a_h)}\,{\rm e}^{2U} - \zeta \right)}
{u^5 + u^3\,(\zeta -1)\,{\rm e}^{-2U(a_h)}\,
{\rm e}^{-4U} - u\,\left({\rm e}^{2U(a_h)}+(\zeta - 1)\,{\rm e}^{-4U(a_h)}\right)\,{\rm e}^{-2U}} .
\end{eqnarray}
For large $\rho$ we have ${\rm e}^{6U} = H \approx 
1+ q/(a_h^2 \, \rho^2)$, and hence we obtain
\begin{eqnarray}\label{FSTU2}
 \frac{F(\rho)}{a_h} = \frac{1}{a} -\frac{\eta}{4\,a^4} \,,
\end{eqnarray}
 where now $\eta=\zeta\,a_h^3=a_h^3+k\,{\rm e}^{2U(a_h)}\,a_h$.

Computing the associated boundary energy-momentum tensor yields
\begin{eqnarray}
\label{eq:tmn-STU2}
16\pi G_5\, \langle T_{\mu\nu} \rangle
&=& \frac{1}{4}\left(R_{\alpha\beta} 
R^{\alpha\;\;\;\beta}_{\,\;\,\mu\;\;\;\nu} -\frac{R^2 }{12}g_{\mu\nu}\right) \\ \nonumber 
&&+ w_5\,M\,\left(g_{\mu\nu}+4\,u_\mu\,u_\nu\right)+\frac{2\,q}{3}\,u_{(\mu}\,R_{\nu)\lambda}\,u^\lambda -2\,\eta \,\sigma_{\mu\nu} \,.
\end{eqnarray}
In units where $L = 16 \pi G_5 =1$, 
the ratio $\eta/s$ reads
\begin{equation}
\frac{\eta}{s} = \frac{1}{4 \pi} \left( 1 + \frac{k \,
{\rm e}^{2U(a_h)}}{a^2_h} \right) \;.
\label{eq:shear-stu-1-ch}
\end{equation}
We note that the correction to $\eta/s = 1/(4 \pi)$ is determined
by the coefficient of the $u \, R \, u$-term in the line 
element \eqref{STUblackPert2}.

\section{Conclusions \label{concl}}

As mentioned in the introduction, 
the energy of a perfect fluid on a three-sphere dual to a static
black hole is not a purely extensive quantity
\cite{Verlinde:2000wg,Savonije:2001nd}.  It contains a subextensive piece $E_c$ which is defined as the violation
of the thermodynamic Euler relation.
In the context of $N=2$ gauged supergravity theories, 
the ratio of $E_c$ and the extensive part $E_e$ of the energy, 
when expressed in terms of black hole data, reads
(in units where $L = 16 \pi G_5 =1$)
\cite{Cardoso:2008gm}
\begin{equation}
\frac{E_c}{E_e} = 
6\, k\, \frac{{\tilde W}_h}{W_h} 
\, \left(\frac{4 \pi}{s} \right)^{2/3} 
 \;,
\label{eq:ratio-EcEe}
\end{equation}
where $s = {\cal S}/vol(S^3)$, and where
${\tilde W}_h$ and $W_h$ denote the superpotentials evaluated
at the horizon.
The Schwarzschild and the Maxwell black hole both satisfy
$W_h = 3 , {\tilde W}_h = 1$.  For these two black holes,
the ratio $\eta/s$ in \eqref{eq:shear-mxw1} can be written as 
\begin{equation}
\frac{\eta}{s} = \frac{1}{4 \pi} \, 
\frac{E_e + \frac12 \, E_c}{E_e} = \frac{1}{4 \pi} \, 
\left( 1 + 3 \, k\, \frac{{\tilde W}_h}{W_h} \,
\left(\frac{4 \pi}{s} \right)^{2/3} 
\right) \;,
\label{eq:eta-s}
\end{equation}
and thus it takes the form \eqref{eq:change-ration-schw}.

The ratio displayed in \eqref{eq:eta-s} takes a form that
is written in manifest $N=2$ language and that could, a priori,
be applicable to any black hole in an $N=2$ model.   However, inspection of the two-charge result \eqref{eq:shear-stu-2} and of the one-charge result \eqref{eq:shear-stu-1-ch} shows that they are not
simply captured by \eqref{eq:eta-s}.  These two cases involve non-trivial scalar fields, and it is 
conceivable that additional terms involving these will have to be added to \eqref{eq:eta-s} in order to 
obtain an expression that is valid for a general $N = 2$ model.

Let us now discuss the diffusion coefficient $D$, defined as in 
\eqref{eq:eta-D}.  Let us first consider the Schwarzschild case, for 
which \eqref{eq:eta-s} implies that the ratio $D = \eta/( \rho + p)=
3 \eta/(4 \rho)$ equals $D = \pi^{1/3}/(4^{2/3} \, s^{1/3})$, as in the black brane case \eqref{eq:eta-s-e}.  Thus, when viewed as a function of $s$, $D$ does not change its functional form. On the
other hand, if $D$ is viewed as a function of the temperature (the
energy), then $D$ will change its functional form due to the subextensive contribution $E_c \propto k$ to the total energy,
i.e. $D$ will not anylonger be simply given in terms of the inverse of the temperature. 
Either way,
$\eta = D \, (\rho + p)$ will receive a correction proportional
to $E_c \propto k$ (see \eqref{eq:ratio-EcEe}).

Next, let us consider the Maxwell case.  Viewing $D$ as a function of $s$, we find that $D$ is not anylonger given by 
$D = \pi^{1/3}/(4^{2/3} \, s^{1/3})$.  This can be understood as
follows.  The total energy of the system is not simply
$E_e + \frac12 \, E_c$, but rather $E_e + \frac12 \, E_c + \frac12 \, Q_A \, \phi^A_h$, where $\phi_h^A$ denote the
electrostatic potentials at the horizon \cite{Cardoso:2008gm}.  
The contribution $Q_A \, \phi^A_h$ is a subextensive contribution
that is distinct from the subleading contribution $E_c$.  The
former is proportional to the square of the charge, while the latter
is proportional to $k$.  Using \eqref{eq:eta-s}, we find that 
the diffusion coefficient $D$ is proportional to the ratio
$(E_e + \frac12 \, E_c) / (E_e + \frac12 \, E_c + \frac12 \, Q_A \, \phi^A_h)$.  At first order in $E_c$, the correction proportional
to $k$ cancels out, while the term proportional to $Q_A \, \phi^A_h$
changes the functional dependence of $D$ on $s$, an effect already observed in \cite{Son:2006em} in the context of charged 
black branes. Thus, when $D$ is viewed as a function of $s$,
it does not receive a correction of order $k$.  However, 
if $D$ is viewed as a function of the temperature (the
energy), then $D$ will change its functional form (at first order in $k$) due to the subextensive contribution $E_c$ to the total energy. Either way,
$\eta = D \, (\rho + p)$ will receive a correction proportional
to $E_c \propto k$.

And finally, in the case of charged black holes with scalar fields,
we find that $D$, when viewed as a function of $s$, receives a
correction of order $k$, since in these cases the term
proportional to $k$ in $\eta$ does not equal $E_c$,
and hence it differs from the
contribution $E_c$ contained in the total energy. Thus, at first order in $k$,
$\eta= D(s) (\rho + p)$ is not any longer given by (1.2).

In deriving the expressions for $\eta/s$ such as (\ref{eq:eta-s}) we restricted ourselves to corrections
of order $k$.  Higher corrections in $k$ are in principle also possible.
For simplicity, 
we took the 
velocity field $u^{\mu}$ of the fluid to be of the specific 
form \eqref{eq:velocity-field}.
Our expressions for $\eta/s$
should, however, be independent of this particular choice
of the velocity field.
We also note that in all the cases considered here,
the deviation from $\eta/s = 1/(4 \pi)$ is determined by the
coefficient of a $u \, R \,u$-term in the associated line element.

In principle, the hydrodynamic energy-momentum tensor may contain 
additional terms, constructed out of derivatives of the velocity field and/or the curvature tensor on the sphere,
that also contribute to $\sigma_{\mu \nu}$
at the same order as the curvature corrections computed in this paper.
However, such terms cannot be present for the solutions constructed here, neither at the order considered in the paper ($\epsilon \, \delta^3$) nor at the next order in derivatives ($\epsilon \, \delta^4$). Such terms would have to be constructed
from the quantities listed in \cite{Bhattacharyya:2008ji} on page 22, which contains a comprehensive
study of the allowed hydrodynamic quantities classified by their tensorial structure.
Since all the quantities appearing in this list either vanish on the solutions considered here
or lead to terms that are of higher order in $\epsilon$, it follows that such terms are absent at order $\epsilon$.

\section*{Acknowledgements}

We would like to thank J. Erdmenger, 
M. Haack, R. Loganayagam, S. Minwalla, S. Nampuri,
G. Policastro, A. Starinets, E. Witten and A. Yarom for valuable
discussions. 
The work of G. Dall'Agata is
partially supported by the Fondazione Cariparo Excellence Grant {\it String-derived supergravities with branes and fluxes and their phenomenological implications} and by the ERC Advanced Grant no. 226455 {\it Supersymmetry, Quantum Gravity and Gauge Fields} (SUPERFIELDS).
The work of G. L. Cardoso is 
supported in part by the Alexander von Humboldt Foundation, by
the Center for Mathematical Analysis, Geometry and Dynamical
Systems (CAMGSD) and by Funda\c{c}\~ao para a Ci\^encia e a Tecnologia (FCT) through the Program Ci\^encia 2008.

\appendix

\section{Very special geometry conventions \label{sec:VSG}}

The five-dimensional $N=2$ gauged supergravity action is based on a set
of real scalar fields $X^A$ that satisfy the constraint
\begin{equation}
\frac16 \, C_{ABC} \, X^A \, X^B \, X^C =1 \;.
\label{vsgconstraint}
\end{equation}
The metric ${\cal G}_{AB}$ is given by
\begin{equation}
{\cal G}_{AB} = - \frac12 \, C_{ABC} \, X^C + \frac92 \, X_A \, X_B \;,
\end{equation}
where
\begin{equation}
X_A = \frac16 \, C_{ABC} \, X^B \, X^C \;.
\end{equation}
Observe that $X^A \, X_A =1$ in view of \eqref{vsgconstraint}.
In addition,
\begin{equation}
X_A \, \partial_i X^A =0 \;,
\label{vsgxdx}
\end{equation}
where $X^A = X^A (\varphi^i)$ and 
$\partial_i X^A (\varphi) = \partial X^A/ \partial \varphi^i$.  Here
the $\varphi^i$ denote the physical scalar fields with target-space
metric
\begin{equation}
{\cal G}_{ij} = {\cal G}_{AB} \, \partial_i X^A \, \partial_j X^B \;.
\end{equation}
A useful relation is 
\begin{equation}
{\cal G}_{AB} \, \partial_i X^B = - 
\frac32 \, \partial_i X_A \;.
\label{gderxx}
\end{equation}
The potential $V_{\rm pot}$ is expressed in terms of 
the superpotential
\begin{equation}
W =  h_A \, X^A 
\label{supo}
\end{equation}
and reads
\begin{equation}
V_{\rm pot} = \mathfrak{g}^2 
\left( {\cal G}^{ij} \, \partial_i W \,\partial_j W - 
\frac43 \, W^2 \right)
= 
\mathfrak{g}^2 \left( h_A \, {\cal G}^{AB} \, h_B - 2 \, W^2 \right)
\;,
\label{potent}
\end{equation}
where in the second step we used
\begin{equation}
{\cal G}^{ij}\, \partial_i X^A 
\,\partial_j X^B = {\cal G}^{AB} - \frac23 \, X^A \, X^B \;.
\label{gijGab}
\end{equation}
The STU model is based on $X^1 X^2 X^3 = 1$, and its metric ${\cal G}_{AB}$ is given by
\begin{equation}
{\cal G}_{AB} = \frac12 \, \delta_{AB} \, \left(X^A \right)^{-2} \;,
\end{equation}
where here there is no summation over $A$.

\section{Rotating Maxwell black hole in Eddington--Finkelstein type
coordinates \label{bl-ef}}

The general non-extremal rotating black hole solution 
in minimal five-dimensional gauged supergravity 
has been given in \cite{Chong:2005hr} 
in Boyer--Lindquist type coordinates.  To 
linear order in angular velocities $\epsilon \, \omega_1$ 
and $\epsilon \, \omega_2$ it reads (with $w_5 = L=k=1$)
\begin{eqnarray}
 ds^2&=&\left(-(1+a^2)+\frac{\Sigma}{a^4}\right)dt^2 + 
\frac{a^2}{\Delta_a}da^2 + a^2 d\theta^2 
+ a^2\left(\sin^2\theta\,d\phi^2 + \cos^2\theta\,d\psi^2\right) \nonumber\\
& &- \frac{2}{a^4}\left(\epsilon\,\omega_2\, \Sigma + \epsilon\,\omega_1\, Q\, 
a^2\right)\cos^2\theta\,d\psi\,dt  
- \frac{2}{a^4}\left(\epsilon\,\omega_1\,\Sigma + 
\epsilon\,\omega_2 \,Q \,a^2\right)\sin^2\theta\, d\phi\,dt  \;,\nonumber\\
A&=&\frac{\sqrt{3}\,Q}{a^2}\left(dt - 
\epsilon \, \omega_1\,\sin^2\theta\,d\phi - 
\epsilon \, \omega_2\,\cos^2\theta\,d\psi\right) \;,
\label{Kerr}
\end{eqnarray}
where
\begin{equation}
\Delta_a= a^2(1+a^2)+\frac{Q^2}{a^2}-M \;\;\;,\;\;\;
\Sigma=M\,a^2-Q^2
\;.
\end{equation}
The line element in \eqref{Kerr} 
can be rewritten in terms of
Eddington--Finkelstein type coordinates by applying the following
transformations,
\begin{equation}
 dt \rightarrow dt-\frac{a^2}{\Delta_a}\,da \;\;\;,\;\;\;
 d\phi \rightarrow d\phi - \frac{\epsilon\,\omega_1}{\Delta_a}
\left(1+a^2\right)da \;\;\;,\;\;\;
 d\psi \rightarrow d\psi - \frac{\epsilon\,\omega_2}{\Delta_a}
\left(1+a^2\right)da \;.
\end{equation}
Then,  to first order in $\epsilon$, the line element becomes
\begin{eqnarray}\label{Kerr2}
 ds^2&=&-\frac{\Delta_a}{a^2}\,dt^2 + a^2\,d\Omega_3^2 + 2\,dt\,da \nonumber\\
& &+\frac{2\,\epsilon}{a^4}\left(\omega_1\,Q^2-\omega_1\,M\,a^2-
\omega_2\,Q\,a^2\right)\,\sin^2\theta\,dt\,d\phi \nonumber\\
& &+\frac{2\,\epsilon}{a^4}\left(\omega_2\,Q^2-\omega_2\,M\,a^2-
\omega_1\,Q\,a^2\right)\,\cos^2\theta\,dt\,d\psi \nonumber\\
& &+2\,\epsilon\,\sin^2\theta\left(\frac{\omega_2\,Q}{\Delta_a}
-\omega_1\right)\,da\,d\phi 
+2\,\epsilon\,\cos^2\theta\left(\frac{\omega_1\,Q}{\Delta_a}
-\omega_2\right)\,da\,d\psi \;,
\end{eqnarray}
while the 
gauge field is still given by \eqref{Kerr}.  Rewriting 
the five-dimensional 
line element \eqref{Kerr2} in terms of the four-dimensional 
quantities $u^\mu=(\,1,\,0,\,\epsilon\,\omega_1,\,\epsilon\,\omega_2)$, $l^\mu=(\,0,\,0,\,-\epsilon\,\omega_2,\,-\epsilon\,\omega_1)$ and $g_{\mu\nu}=diag(\,-1,\,1,\,\sin^2\theta,\,\cos^2\theta)$ yields the line element 
(\ref{line-maxdef-1})
with $\sigma_{\mu \nu} =0$ and $\kappa = -1/(2 \sqrt{3})$.

\section{A two-charge rotating STU black hole in Eddington--Finkelstein type coordinates
\label{STUblackhole}}

A rotating version of the static two-charge 
STU black hole solution (\ref{staticfields}) 
has been constructed in \cite{Chong:2005da}. 
To linear order in rotation parameters $\epsilon\,\omega_1$ and $\epsilon\,\omega_2$ it reads
\begin{eqnarray}\label{STUblack}
 ds^2 &=& H^{-\frac43}\,\left[\,-\frac{X}{r^2}\,dt^2 + \frac{2\,\epsilon}{r^2}\,\left(X-\frac{f_3}{r^2}\right)\,\left(\omega_1\,\sin^2\theta\,dt\,d\phi + \omega_2\,\cos^2\theta\,dt\,d\psi\right)\right.   \nonumber \\
& & +\left.\,\frac{f_3^2}{r^6}\,\left(\sin^2\theta\,d\phi^2 + \cos^2\theta\,d\psi^2\right)\right] +H^{\frac23}\,\left[\,\frac{r^2}{X}\,dr^2 + r^2\,d\theta^2\right] \,, \nonumber \\
H &=& 1+\frac{\mu\,s^2}{r^2} \,, \nonumber \\
X &=& r^2-\mu+\mathfrak{g}^2\,(r^2+\mu\,s^2)^2  \,, \nonumber \\
f_3 &=& r^4+ \mu\,s^2\,r^2 \,, 
\end{eqnarray}
where
\begin{equation}
 s=\sinh\,\delta \,, \qquad c=\cosh\,\delta \,.
\end{equation}
(Here, $s$ should not be confused with the entropy density
in the main text.)
The associated gauge potentials are
\begin{eqnarray}\label{gfields}
 A^1 &=& A^2 = \frac{\mu\,s\,c}{r^2 H}\,\left(dt-\epsilon\,(\,\omega_1\,\sin^2\theta\,d\phi + \omega_2\,\cos^2\theta\,d\psi)\right) \,,\nonumber \\
A^3 &=& \frac{\mu\,s^2}{r^2}\,\epsilon\,\left(\omega_2\,\sin^2\theta\,d\phi + \omega_1\,\cos^2\theta\,d\psi\right) \,.
\end{eqnarray}
Setting $\mathfrak{g}^2=1$, $\mu\,s^2=q$, $\mu\,s\,c=Q$ as well as
changing the radial coordinate to $a=r\,H^{\frac13}$
and carrying out the transformations
\begin{eqnarray}
 dt&\rightarrow&dt - \frac{3\,a^2\,H^{\frac23}}{X\,W(a)}\,da \,,
 \\
 d\phi&\rightarrow&d\phi+\epsilon\,\omega_1\left(dt - \frac{3\,H^{-\frac13}}{a^4 f(a)\,W(a)}\,da\right)\,,\quad  d\psi \rightarrow d\psi+\epsilon\,\omega_2\left(dt - \frac{3\,H^{-\frac13}}{a^4 f(a)\,W(a)}\,da\right)\,, \nonumber
\end{eqnarray}
where $f(a)$ is given in (\ref{eq:function-f}) with $k$ set 
to $k=1$ and with ${\rm e}^{3U} = H$,
yields \eqref{STUblack} and \eqref{gfields} in Eddington--Finkelstein type coordinates to first order in $\epsilon$,
\begin{eqnarray}\label{STUblackEdd}
ds^2 &=&-a^2\,f(a)\,dt^2 +\frac{6}{W(a)}\,da\,dt - \frac{6}{W(a)}\,\epsilon\,\left(\omega_1\,\sin^2\theta\,d\phi+\omega_2\cos^2\theta\,d\psi\right)\,da \nonumber \\
& & +a^2\,d\Omega_3^2 + 2\,\epsilon\,\left(a^2\,f(a)+a^2-H^{-\frac13}\right)\left(\omega_1 \sin^2\theta\,d\phi+\omega_2\cos^2\theta\,d\psi\right)\,dt \,, \nonumber \\
A^1 &=& A^2 = -\frac{Q}{a^2}\,H^{-\frac13}\,u_\mu\,dx^\mu \,,\quad A^3=-\frac{q}{a^2}\,H^{\frac23}\,l_\mu\,dx^\mu \,,
\end{eqnarray}
where $W(a)=2\,H^{-\frac{1}{3}}+H^{\frac23}$ is the superpotential. Then, rewriting the five-dimensional line element \eqref{STUblackEdd} and gauge potentials \eqref{gfields} in terms of the four-dimensional quantities $u^\mu=(\,1,\,0,\,\epsilon\,\omega_1,\,\epsilon\,\omega_2)$,  $l^\mu=(\,0,\,0,\,-\epsilon\,\omega_2,\,-\epsilon\,\omega_1)$ 
and $g_{\mu\nu}=diag(\,-1,\,1,\,\sin^2\theta,\,\cos^2\theta)$ yields \eqref{STUblackPert} with $\sigma_{\mu \nu} = 0$.

\section{Three-charge rotating STU black hole with equal rotation parameters in Eddington--Finkelstein coordinates
\label{genSTU}}

A rotating three-charge STU black hole with equal rotation
parameters $\omega_1 = \omega_2 = \tilde \omega$ has been constructed
in \cite{Cvetic:2004ny}.  
To first oder in the rotation parameter $\epsilon \,
\tilde \omega$ it reads
\begin{eqnarray}\label{STUblackgen}
 ds^2 &=& -\frac{Y}{R^2}\,dt^2 + \frac{r^2\,R}{Y}\,dr^2 + R\,d\Omega_3^2 - \frac{2\,f_2}{R^2}\,dt\,\left(\sin^2\theta\,d\phi + \cos^2\theta\,d\psi \right) \,, \nonumber \\
A^i &=& \frac{\mu}{r^2\,H_i}\,\left(s_i\,c_i\,dt + \epsilon\,\tilde{\omega}\,\left(c_i\,s_j\,s_k - s_i\,c_j\,c_k\right)\left(\sin^2\theta\,d\phi + \cos^2\theta\,d\psi\right)\right) \,, \nonumber \\
X^i &=& \frac{R}{r^2\,H_i} \,,\quad i=1,\,2,\,3 \,,\quad i\neq j\neq k \neq i \, ,
\end{eqnarray}
where
\begin{eqnarray}
Y &=& R^3 + r^4 -\mu\,r^2 \,, \nonumber \\
R &=& r^2\,\left(\prod_{i=1}^3\,H_i\right)^\frac13 \,,\qquad H_i = 1+\frac{\mu\,s_i^2}{r^2} \,,\nonumber \\
f_2 &=& \epsilon\,\tilde{\omega}\,\left(-\gamma\,R^3 + \mu\,\left(\prod_{i}\,c_i - \prod_{i}\,s_i\right)\,r^2 + \mu^2\,\prod_{i}\,s_i\right) \,,\nonumber \\
s_i &=& \sinh\delta_i \,,\qquad c_i = \cosh\delta_i \,.
\end{eqnarray}
Changing the radial coordinate to $a=r\,{\rm e}^{U}=r\,\left(H_1\,H_2\,H_3\right)^{\frac16}$ and applying the transformation
\begin{equation}
dt \rightarrow dt -\frac{3}{W(a)\,a^2\,f(a)}\,da \,, 
\end{equation}
yields the line element \eqref{STUblackgen} in the form
\begin{eqnarray}
 ds^2 &=& -a^2\,f(a)\,dt^2 + \frac{6}{W(a)}\left(dt + \frac{\epsilon\,\tilde{\omega}\,h(a)}{a^2\,f}\,\left(\sin^2\theta\,d\phi + \cos^2\theta\,d\psi\right) \right)\,da \nonumber \\
& &+a^2\,d\Omega_3^2 - 2\,\epsilon\,{\tilde \omega}\,h(a)\,\left(\sin^2\theta\,d\phi + \cos^2\theta\,d\psi\right)\,dt \,,
\end{eqnarray}
where
\begin{eqnarray}\label{STUblackgen2}
 h(a) = -\gamma\,a^2 +\frac{\mu}{a^2}\,{\rm e}^{-2U}\,\left(\prod_{i}\,c_i - \prod_{i}\,s_i\right) + \frac{\mu^2}{a^4}\,\prod_{i}\,s_i  \,.
\end{eqnarray}
For later convenience we define $\omega=\gamma\,\tilde{\omega}$ and $\tilde{h}=\gamma^{-1}\,h$ such that $\omega\,\tilde{h}=\tilde{\omega}\,h$. Then carrying out the transformations
\begin{equation}
 d\phi \rightarrow d\phi + \epsilon\,\omega \left(dt - \frac{3\left(a^2 f(a)+\tilde{h}(a)\right)}{a^4\,f(a)\,W(a)}\,da\right), d\psi \rightarrow d\psi + \epsilon\,\omega \left(dt- \frac{3\left(a^2 f(a)+\tilde{h}(a)\right)}{a^4\,f(a)\,W(a)}\,da\right) 
\end{equation}
yields \eqref{STUblackgen} in Eddington--Finkelstein type
coordinates to first order in $\epsilon$,
\begin{eqnarray}\label{STUblackgen3}
ds^2 &=& -a^2\,f(a)\,dt^2 + \frac{6}{W(a)}\left(dt - \epsilon\,\omega\left(\sin^2\theta\,d\phi + \cos^2\theta\,d\psi\right)\right)\,da  \nonumber \\
& & + a^2\,d\Omega_3^2 + 2\,\epsilon\,\omega\left(a^2\,f(a) + a^2 - \left(a^2\,f(a)+\tilde{h}(a)\right)\right)\left(\sin^2\theta\,d\phi + \cos^2\theta\,d\psi\right)\,dt\,, \nonumber \\
A^i &=& \frac{\mu}{a^2\,H_i}\,e^{2U}\left(s_i\,c_i\,dt + \epsilon\,\frac{\omega}{\gamma}\,\left(c_i\,s_j\,s_k - s_i\,c_j\,c_k\right)\left(\sin^2\theta\,d\phi + \cos^2\theta\,d\psi\right)\right) \,, \nonumber \\
X^i &=& \frac{1}{H_i}\,\left(\prod_{i=1}^3\,H_i\right)^\frac13 \,,\quad i=1,\,2,\,3 \,,\quad i\neq j\neq k \neq i\,.
\end{eqnarray}

The line element in (\ref{STUblackgen3}) is related to
the various line elements used in the main text, as follows.
Let us first consider the stationary limit of the Maxwell
solution \eqref{line-maxdef-1} with $\omega_1=\omega_2=\omega$.
It is obtained from 
\eqref{STUblackgen3} by setting $\delta_1=\delta_2=\delta_3=\delta$, $W(a)=3$ and $\Delta_a = f a^4$ with $f$ given by $f(a) = 1 + k/a^2 - M/a^4 + Q^2/a^6 $. Then the function $h$ becomes
(with $s_i = s, c_i = c$)
\begin{eqnarray}
h(a)=-\gamma\,a^2 + \frac{\mu}{a^2}\,{\rm e}^{-2U}\left(c^3-s^3\right)+\frac{\mu^2}{a^4}\,s^3 \;,
\end{eqnarray}
which can also be written as
\begin{eqnarray}
h(a)= \gamma\,\tilde{h}(a)=(c-s)\left(-a^2  + \frac{\mu}{a^2}\,{\rm e}^{-2U}\,(c^2+s^2+c\,s)+\frac{\mu^2}{a^4}\,s^3\,(c+s)\right) \,.
\end{eqnarray}
Setting $M=\mu+2\,\mu\,s^2$, $Q=\mu\,s\,c$ and 
${\rm e}^{-2U}=(a^2-\mu\,s^2)/a^2$ gives
\begin{eqnarray}
\tilde{h}(a)=-a^2\,f(a) + 1 + \frac{Q}{a^2} \,.
\end{eqnarray}
The terms in this expression are related as follows to the ones
in \eqref{line-maxdef-1}:
the second term is the coefficient of the $u\,R\,u$-term, while the third term is the coefficient of the $u\, l$-term.

Next, let us consider the stationary limit of the two-charge
solution \eqref{STUblackPert}. It is obtained from 
\eqref{STUblackgen3}
by setting $\delta_1=\delta_2=\delta$, $\delta_3=0$, $\gamma=1$ and $H={\rm e}^{3U}$. Then the function $h$ becomes
\begin{eqnarray}
h(a)=\tilde h(a)=-a^2\,f(a) +{\rm e}^{-U} \,,
\end{eqnarray}
with $f$ given by \eqref{eq:function-f}.  In this expression, 
the second term is the coefficient of the $u\,R\,u$-term in \eqref{STUblackPert}. 

And finally, the stationary limit of the one-charge
solution \eqref{STUblackPert2} is obtained from 
\eqref{STUblackgen3}
by setting
$\delta_1=\delta$ and $\delta_2=\delta_3=0$.  Now the 
function $h$ reads (with $c_1 = c$)
\begin{eqnarray}
 h(a)=-\gamma\,a^2 + \frac{\mu}{a^2}\,{\rm e}^{-2U}\,c \,.
\end{eqnarray}
This can be written as 
\begin{eqnarray}
 h(a)=\gamma\,\tilde{h}(a)=\frac1c\left(-a^2 + \frac{\mu}{a^2}\,{\rm e}^{-2U}\,c^2\right) \,.
\end{eqnarray}
Setting $\gamma=c^{-1}$, and with $H={\rm e}^{6U}$, we obtain
\begin{eqnarray}
\tilde{h}(a)=-a^2\,f(a) + {\rm e}^{2U} \,,
\end{eqnarray}
with $f$ given by \eqref{eq:function-f}.  In this expression, 
the second term is the coefficient of the $u\,R\,u$-term in \eqref{STUblackPert2}.

\section{Boundary energy-momentum tensor for the STU black hole solution 
(\ref{STUblackPert})
\label{tmnSTU}}

Here we compute the boundary energy-momentum for
the STU black hole carrying two equal charges. A similar
calculation applies to the other cases discussed in the main text,
namely no charge (the Schwarzschild case), one non-vanishing charge and  three equal charges
(the Maxwell case).

The boundary energy-momentum 
tensor is given by 
\cite{Balasubramanian:1999re,Emparan:1999pm,Batrachenko:2004fd}
\begin{equation}
8 \pi G_5 \, \langle T_{\mu \nu} \rangle = \lim_{a \rightarrow \infty} \left[a^2 \left(
K_{\mu \nu} - K \, \gamma_{\mu \nu} - \frac{W(a)}{L} \,\gamma_{\mu \nu} 
+ \frac{L}{2} \, G_{\mu \nu} \right) \right] \;,
\label{TmunuSTU}
\end{equation}
where the boundary metric $\gamma_{\mu \nu}$ is read off from
the bulk metric written in the form
\begin{equation}
ds^2 = N^2 da^2 + \gamma_{\mu \nu} \left( dx^{\mu} + n^{\mu} da \right)
 \left( dx^{\nu} + n^{\nu} da \right) \;,
\label{eq:adm-decomp}
\end{equation}
$G_{\mu \nu} = R_{\mu \nu}[\gamma] - \frac12 \gamma_{\mu \nu} \, R[\gamma]$ 
is the 
four-dimensional Einstein
tensor of $\gamma_{\mu \nu}$, and the extrinsic curvature tensor is given by
\cite{Haack:2008cp}
\begin{equation}
K_{\mu\nu}=-
\frac{1}{2N}\left(\partial_a \gamma_{\mu\nu}-\nabla_\mu[\gamma] \,n_\nu
-\nabla_\nu[\gamma] \, n_\mu\right) \;,
\label{eq:extr-curv}
\end{equation}
with $K=\gamma^{\mu\nu} \, K_{\mu\nu}$.
Here $n_{\mu} = \gamma_{\mu \nu} \, n^{\nu} $, and $W(a)$ is the superpotential.

Imposing the tracelessness of $T_{\mu \nu}$ results in 
$K = - 4\,W(a)/(3\,L) - L\, R[\gamma]/6$, 
and reinserting this into (\ref{TmunuSTU}) yields
\begin{equation}
8 \pi G_5 \, \langle T_{\mu \nu} \rangle = \lim_{a \rightarrow \infty} \left[a^2 \left(
K_{\mu \nu} + \frac{W(a)}{3\,L} \,\gamma_{\mu \nu} 
+ \frac{L}{2} \left(R_{\mu \nu}[\gamma] - \frac16 \, \gamma_{\mu \nu} \, 
R[\gamma] 
\right)  \right) \right] \;.
\label{eq:stress-energy-tensorSTU}
\end{equation}
In the following we set $L=1$.

Comparing (\ref{eq:adm-decomp}) with the line element 
(\ref{STUblackPert}) for the deformed STU black hole, and using 
\eqref{FSTU}, we infer that
for large $a$,
\begin{eqnarray}
n_\mu&=&-\frac{3}{W(a)}\,u_\mu \;\;\;,\;\;\;
N^2=-\frac{9}{W(a)^2}\,\gamma^{\mu\nu}\,u_\mu\,u_\nu\;,\nonumber\\
\gamma_{\mu\nu}&=& a^2 g_{\mu \nu} - \left({\rm e}^{-4U}\,k- {\rm e}^{-2U}\,\frac{\mu}{a^2}
\right)\, u_{\mu} \, u_{\nu} 
+ \frac{1}{2}\,{\rm e}^{-U}\left(u_\mu \, R_{\nu\lambda}\, u^\lambda + u_\nu \, R_{\mu\lambda}\, u^\lambda\right)
\\ \nonumber
& &+ \left(2a-\frac{\eta}{2a^2}\right)\,\sigma_{\mu \nu}\;. 
\end{eqnarray}
Here $W(a)\approx 3 + q^2/(3\,a^4)$ and the exponential functions ${\rm e}^{-\chi\,U}$ in $\gamma_{\mu\nu}$ behave as ${\rm e}^{-\chi\,U} \approx 1-\chi\,q/(3\,a^2)$ so that
\begin{eqnarray}
\gamma_{\mu\nu}&=& a^2 g_{\mu \nu} - \left(k- \frac{w_5\,M}{a^2}
\right)\, u_{\mu} \, u_{\nu} 
+ \frac{1}{2}\left(1-\frac{q}{3\,a^2}\right)\left(u_\mu \, R_{\nu\lambda}\, u^\lambda + u_\nu \, R_{\mu\lambda}\, u^\lambda\right)
\\ \nonumber
& &+ \left(2a-\frac{\eta}{2a^2}\right)\,\sigma_{\mu \nu}\;,
\end{eqnarray}
where $w_5\,M=\mu + \frac43\,k\,q$ is the physical mass.
At first order in $\epsilon$ and at large $a$, 
the inverse metric $\gamma^{\mu\nu}$ is then given by
\begin{eqnarray}
\gamma^{\mu\nu}=\frac{1}{a^2}\,g^{\mu\nu}+\frac{k}{a^4}
\,u^\mu\,u^\nu
-\frac{1}{2 a^4}
\left(u^\mu \, R^\nu_{\;\lambda}\, u^\lambda + u^\nu \, R^\mu_{\;\lambda}\, u^\lambda\right)
-\frac{2}{a^3}
\,\sigma^{\mu\nu} \;,
\end{eqnarray}
where the indices on the right hand side are raised with the metric $g^{\mu \nu}$.

Computing the terms in (\ref{eq:stress-energy-tensorSTU})
for large $a$ and to first order in $\epsilon$, we obtain 
\begin{eqnarray}\label{stressenexprSTU}
W(a) &=& 3 + \frac{q^2}{3\,a^4} \;, \nonumber \\
N^{-1}&=&a+\frac{k}{2a}-\frac{k^2}{8a^3}-\frac{w_5\,M}{2 a^3}+\frac{q^2}{9a^3} \;,\nonumber\\
-\frac{1}{2N}\partial_a\gamma_{\mu\nu}+\frac{W(a)}{3}\,\gamma_{\mu\nu}&=&\frac{1}{2a^2}\left(\frac{k^2}{4}g_{\mu\nu}+w_5\,M\left(g_{\mu\nu}+4\,u_\mu\,u_\nu\right)-2\,\eta\,\sigma_{\mu\nu}\right) \nonumber\\
& &-\frac{k}{2}\left(g_{\mu\nu}+2\,u_\mu\,u_\nu\right)+\left(a-\frac{k}{2a}\right)\sigma_{\mu\nu}  \nonumber\\
& &+\left(\frac{1}{2}-\frac{q}{3\,a^2}\right)\left(u_\mu \, R_{\nu\lambda}\, u^\lambda + u_\nu \, R_{\mu\lambda}\, u^\lambda\right) \;,\nonumber\\
\Gamma^\gamma_{\alpha\beta}[\gamma]&=&\Gamma^\gamma_{\alpha\beta} +\frac{1}{a}g^{\gamma\lambda}\left(\nabla_\alpha\sigma_{\lambda\beta}
+\nabla_\beta\sigma_{\alpha\lambda}-\nabla_\lambda\sigma_{\alpha\beta}\right) 
\;,\nonumber\\
\nabla_\mu[\gamma]\,n_\nu&=&-\nabla_\mu\,u_\nu \;,\nonumber\\
\label{Rmunu}R_{\mu\nu}[\gamma]&=&R_{\mu\nu} +\frac{4k}{a}\,\sigma_{\mu\nu} 
\;,\\ 
R[\gamma]&=&\frac{R }{a^2} \;, \nonumber\\
R_{\mu\nu}[\gamma]-\frac16\gamma_{\mu\nu}R[\gamma]&=&R_{\mu\nu} -k g_{\mu\nu}+\frac{k^2}{a^2}\,u_\mu\,u_\nu 
-\frac{k}{2a^2}\left(u_\mu \, R_{\nu\lambda}\, u^\lambda + u_\nu \, R_{\mu\lambda}\, u^\lambda\right)
+\frac{2k}{a}\sigma_{\mu\nu}
\;.\nonumber
\end{eqnarray}
Inserting these expressions into 
(\ref{eq:stress-energy-tensorSTU}) yields
the energy-momentum tensor (\ref{eq:tmn-STU}).


\providecommand{\href}[2]{#2}
\begingroup\raggedright\endgroup

\end{document}